\documentclass[10pt]{article}
\usepackage[utf8]{inputenc}

\usepackage[top=2.5cm, bottom=2.5cm, left=2cm, right=2cm]{geometry}

\usepackage[british]{babel}
\usepackage[T1]{fontenc}
\usepackage{times}
\usepackage{makecell, cellspace}
\usepackage{amsmath}
\usepackage{titling}
\usepackage[font=small,labelfont=bf,calcwidth=\linewidth]{caption}
\usepackage{amsfonts}
\usepackage{mathrsfs}
\usepackage[hidelinks]{hyperref}
\usepackage{mathtools}
\usepackage{tikz}
\usetikzlibrary{decorations.pathreplacing, calligraphy}
\usepackage{mathtools}
\usepackage{multirow,tabularx}
\usepackage{xfrac}
\usepackage{enumitem}
\usepackage{natbib}
\usepackage{booktabs}
\usepackage{listings}
\usepackage{verbatim}
\usepackage{printlen}
\usepackage{xcolor}
\AtBeginDocument{\colorlet{defaultcolor}{.}}

\usepackage{hhline}
\usepackage[section]{algorithm}
\usepackage{algorithmic}
\usepackage{nicefrac}
\usepackage[textsize=scriptsize]{todonotes}
\usepackage{comment}
\usepackage{sectsty}
\usepackage{float}
\usepackage{placeins}
\usepackage{flafter}
\usepackage[htt]{hyphenat}

\usepackage{microtype}

\DeclareMathAlphabet{\mathbbold}{U}{bbold}{m}{n}

\allsectionsfont{\sffamily}

\newcommand*{\veps}{\varepsilon}

\let\sPhi\Phi
\renewcommand*{\Phi}{\varPhi}

\newcommand*{\dd}{\ensuremath{\mathrm{d}}}
\newcommand*{\dx}[1]{\ensuremath{\,\dd{#1}}}

\newcommand*{\deltax}{\ensuremath{{\delta}x}}
\newcommand*{\deltau}{\ensuremath{{\delta}u}}

\newcommand*{\RR}{\ensuremath{\mathbb{R}}}

\DeclareMathOperator{\tr}{tr}

\newcommand*{\abs}[2][{}]{\ensuremath{#1\left\lvert#2#1\right\rvert}}

\newcommand*{\bfj}{\ensuremath{\mathbf{j}}}
\newcommand*{\bfB}{\ensuremath{\mathbf{B}}}
\newcommand*{\bfE}{\ensuremath{\mathbf{E}}}

\newcommand*{\bfe}{\ensuremath{\mathbf{e}}}

\newcommand*{\dlst}{\ensuremath{\Delta^{\!\ast}}}

 \let\epsilon\veps
 
\setuptodonotes{inline}

\makeatletter
\def\@seccntformat#1{\csname the#1\endcsname.\quad}
\makeatother


\begin{document}

{\vspace*{2cm}
\sffamily\huge\raggedright\bfseries FGE: A Fast Free-Boundary Grad-Shafranov Evolutive Solver\par
\normalfont\normalsize
\vspace*{1cm}

\begin{minipage}{0.9\textwidth}
\large\sffamily\raggedright\bfseries
Cosmas~Hei\ss{}\textsuperscript{\normalfont1}, Antoine~Merle\textsuperscript{\normalfont1}, Francesco~Carpanese\textsuperscript{\normalfont1}, Federico~Felici\textsuperscript{\normalfont 3,1}, Craig~Donner\textsuperscript{\normalfont 3}, Stefano~Marchioni\textsuperscript{\normalfont 1,}\footnotemark[1], Alessandro~Mari\textsuperscript{\normalfont 2}, Olivier~Sauter\textsuperscript{\normalfont 1}\par
\vspace*{0.5cm}
\normalfont\small
\textsuperscript{1} \'Ecole Polytechnique F\'ed\'erale de Lausanne (EPFL), Swiss Plasma Center (SPC), Lausanne, Switzerland\\
\textsuperscript{2} \'Ecole Polytechnique F\'ed\'erale de Lausanne (EPFL), Swiss Data Science Center (SDSC), Lausanne, Switzerland\\
\textsuperscript{3} Google DeepMind, London, United Kingdom
\end{minipage}
\vspace*{1cm}

\begin{minipage}{0.7\textwidth}
\normalfont\normalsize\sffamily\raggedright\bfseries Abstract\par
\vspace*{0.5em}
\normalfont\normalsize\raggedright
Accurate and rapid simulation of the free boundary tokamak plasma equilibrium evolution is essential for modern plasma control, stability analysis, and scenario development. This paper presents the Free-Boundary Grad-Shafranov Evolutive (FGE) code, a highly flexible and control-oriented solver designed to address the challenges posed by advanced plasma configurations across a range of devices.
FGE evolved from the FBT \citep{FBT_code} and LIUQE \citep{liuqe_paper} codes and is part of the MEQ suite, sharing many of the low-level optimized functions.
It self-consistently solves the free-boundary Grad-Shafranov equation coupled with circuit equations for external conductors and models for plasma profile evolution. The code implements a fully non-linear, Newton-based framework with multiple, highly optimized solver options, state representations, and residual formulations that enable rapid computation
across different simulation setups. A key capability is the self-consistent integration of various 0D and 1D current diffusion equations (CDEs) to model the resistive evolution of the plasma current profile as well as the ability to model plasmas with multiple magnetic axes (Doublets). Furthermore, FGE allows to linearize the plasma dynamics around a given equilibrium to generate a state-space model suitable for controller design and analysis. Numerical studies are presented demonstrating the code's speed of convergence and validating its performance against experimental data. Furthermore, benchmarks against the RAPTOR and KINX codes for profile evolution and vertical growth rate estimates are presented highlighting FGE's capabilities for both prediction and analysis.
\end{minipage}}
\footnotetext[1]{Current address: Institute of Plasma Physics of the CAS, Prague, Czech Republic}

\twocolumn

\section{Introduction}
On resistive timescales, tokamak plasmas generally exist in a state of force balance governed by the Grad-Shafranov equation \citep{grad_equi,shafranov_equi}. This equilibrium typically evolves with changes in external conductors and the internal plasma current distribution. Our focus lies in solving for the evolution of the free-boundary Grad-Shafranov equilibrium together with external currents based on voltage inputs to the active coils. The ability to rapidly simulate the plasma response to external actuators finds applications in many domains of tokamak operation, including shot preparation, scenario development and post-shot analysis. It is of particular importance when designing feedback magnetic control strategies as a free-boundary plasma equilibrium evolution model not only provides information about vertical stability and plasma shape response but also enables testing controllers in the loop ahead of deployment.

Consequently, numerous codes have been developed to tackle the free-boundary magnetic equilibrium evolution problem. Early developments of equilibrium evolution codes include the SCED code \citep{BLUM1981235}, the PROTEUS code \citep{albanese1987proteus}, the MAXFEA code \citep{barabaschi1993maxfea} and the Tokamak Simulation Code \citep{JARDIN1986481_TSC}. Among the more modern codes, the list includes SPIDER, \citep{spidercode}, CREATE-NL \citep{create_nl_paper}, FEEQS \citep{cedres_feeqs_paper}, GSevolve \citep{gsevolve_paper}, NICE \citep{nice_paper, GROS2025113849_nice_with_weakCDE}, DINA \citep{DINA_paper}, FRIDA-TD \citep{frida}, Tokamaker \citep{tokamaker} and FEQIS \citep{feqis_paper}. These codes differ in various aspects including their choice of the computational domain, treatment of boundary terms, usage of finite elements or finite differences, solver strategies and the possible addition of 1D plasma profile equations.

This paper presents the Forward Grad-Shafranov Evolutive (FGE) code, an evolution of the work of \citep{carpaneses_thesis}, which also served as a basis for FreeGSNKE \citep{freegsnake}. FGE is an essential component of the Matlab EQuilibrium Suite (MEQ), designed to rapidly solve for the free-boundary Grad-Shafranov equilibrium while also solving for the evolution of coil and vessel currents as well as the resistive evolution of plasma current profiles.
Since its addition to the MEQ suite, FGE has been continuously maintained and improved and has already proven essential in various control-oriented projects. We refer to \citep{degrave_felici_nature, tenaglia_allocator, Mele_shape_control25} for examples of recent works that use FGE for advancing the design of magnetic controllers at TCV. Furthermore, FGE is already tried and tested on different Tokamaks as e.g. SPARC \citep{sparc_abstract_using_fge, wai25}, MAST-U and STEP \citep{mast_abstract_using_fge}. As part of MEQ, FGE is fully integrated with the LIUQE \citep{liuqe_paper} equilibrium reconstruction code and a matlab re-implementation of the FBT \citep{FBT_code} inverse Grad-Shafranov code, enabling streamlined workflows for running free-boundary simulations on the basis of equilibrium reconstructions or on the equilibria computed for shot design. Moreover, while FGE is mainly written in Matlab enabling rapid visualization and interactive use in e.g. analysis and simulation scripts, core functionalities shared across MEQ are implemented in pre-compiled C-functions, significantly improving computational performance.

In its current form, FGE implements a fully non-linear Newton-based approach to efficiently determine the plasma equilibrium. Building on the use of the Jacobian-Free Newton-Krylov solver in \citep{carpaneses_thesis}, FGE was extended to implement fully analytical jacobian computations and offer multiple possible residual formulations enabling new solver strategies and optimizations. FGE offers the flexibility to simulate plasmas with advanced divertor configurations and even multiple magnetic axes like Droplets and Doublets. Optimizations in the subroutines enable FGE to solve a single time step on the order of milliseconds with the standard TCV computational grid on consumer hardware. 
Hence, feedback controllers can be rapidly tested and validated with FGE in the loop and controller settings efficiently optimized. Similarly to NICE \citep{GROS2025113849_nice_with_weakCDE}, based on the weak formulation introduced in the work of \citep{heumann_weak_formulation}, FGE was extended to offer the capability to self-consistently solve a 1D current diffusion equation alongside the Grad-Shafranov equilibrium.

The structure of this paper is as follows: Section~\ref{sec:problem_formulation} presents the system of equations solved in FGE. In section~\ref{sec:solving_the_problem}, we lay out the various methods implemented in the code for solving the nonlinear system using Newton methods. Section~\ref{sec:linearization} outlines the capability of FGE to capture plasma dynamics in a linearized state-space system.
Finally, Section~\ref{sec:results} offers comprehensive results, including benchmarks on different plasma scenarios, convergence speeds, and validation of the current diffusion equation against analytical examples and the RAPTOR code.

\section{Problem Formulation}
\label{sec:problem_formulation}
We consider the problem of solving for the evolution of the free-boundary axisymmetric equilibrium on resistive timescales. We assume these to be much longer than the Alfvén timescales and consequently assume MHD force balance to be fulfilled at each time step. In this case, the plasma evolution is driven by changing currents in active coils, passive conductors and the plasma itself. 

\subsection{Equilibrium Equations}
\label{sec:equilibrium_equations}
In axisymmetric geometry, we adopt a right-handed cylindrical coordinate system $(R, \phi, Z)$ with the COCOS=17 coordinate convention \citep{cocos}. The magnetic field is decomposed in terms of the poloidal flux $\psi$ and the toroidal field component $T/R$:
\begin{equation}
    \bfB = -\frac{1}{2 \pi R} \frac{\partial \psi}{\partial Z} \bfe_R + \frac{1}{2 \pi R} \frac{\partial \psi}{\partial R} \bfe_Z + \frac{T}{R} \bfe_\phi.
\end{equation}
Assuming MHD force balance $\bfj \times \bfB = \nabla p$ for an isotropic pressure tensor yields the Grad-Shafranov equation \citep{grad_equi, shafranov_equi}:
\begin{equation}
\label{eq:GS}
    \dlst \psi = - 2 \pi R \mu_0 j_{\phi} = -4 \pi^2 \left(\mu_0 R^2 p'(\psi) + TT'(\psi)\right),
\end{equation}
expressing the toroidal current density $j_\phi$ inside the plasma domain in terms of the flux functions $p'$ and $TT'$.

In FGE, the Grad-Shafranov equation is discretized on a uniform rectangular grid in $(R,Z)$ containing the limiter perimeter of the machine. The $\dlst$~operator is given by a finite difference representation and the toroidal current density $j_\phi$ is discretized as a collection of ideal current filaments centered at the grid points. We denote by $I_y$ the vector of these currents.

\begin{figure}[t]
    \centering
    \includegraphics[width=0.9\linewidth]{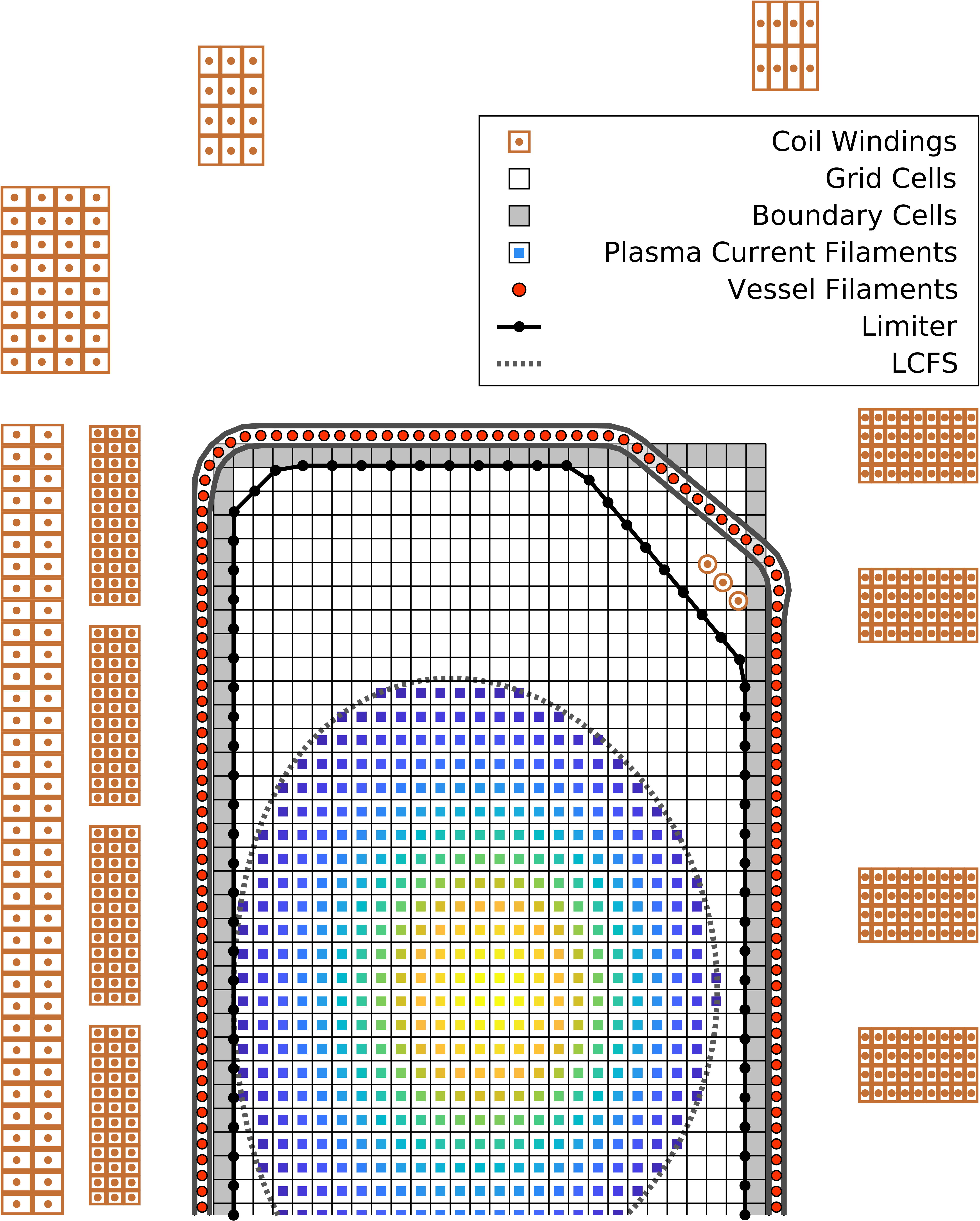}
    \caption{Illustration of the discretization used in FGE showing the computational grid, modeled coil windings, vessel filaments, and the limiter for the top half of TCV. Boundary elements of the grid are highlighted in gray. The plasma of TCV \#61400 at 0.1s is shown in terms of discretized plasma current filaments on the inner grid with the plasma boundary indicated by the dashed contour.}
    \label{fig:discretization}
\end{figure}

The plasma domain is identified by selecting possible limiting points and X-points on the grid and finding the according flux value of the last closed flux surface (LCFS). Magnetic axes are found in a similar way and are used to compute the normalized poloidal flux map. The plasma current filaments $I_y$ are constrained to be zero outside the LCFS. Figure~\ref{fig:discretization} illustrates the computational domain used in FGE. We refer to \citep{liuqe_paper} for a more in-depth description of the discretization and the plasma domain identification.

The poloidal flux at the grid boundary is given by the contribution of plasma currents and coil currents
\begin{equation}
    \psi_b = M_{by} I_y + M_{be} I_e,
\end{equation}
where $M_{\ast \ast}$ are the mutual inductances between concentric conductors given by an analytical expression and $I_e$ is the vector representing external currents, comprised of active currents inside poloidal field coils and passive vessel currents. We note that to speed up computations, the contribution of the plasma current to the boundary flux is usually computed using a method devised by Lackner \citep{lackners_trick}, inverting $\dlst$ with a homogeneous Dirichlet boundary condition and subsequently evaluating the normal derivatives at the domain's boundary.

External coils are discretized as collections of ideal current filaments for the coil windings. Vessel currents are by default discretized into individual current filaments as well, although they can alternatively be represented using combinations thereof, ordered into vessel response eigenmodes, allowing rapidly decaying higher-order modes to be neglected for faster computations. The temporal evolution of the external currents $I_e$ is modeled by the circuit equation
\begin{equation}
\label{eq:circuit}
\begin{split}
    \begin{pmatrix}
        V_a\\
        0
    \end{pmatrix}
    = M_{ee} \dot{I}_e + M_{ey} \dot{I}_y + R_e I_e,
\end{split}
\end{equation}
where $R_e$ are conductor resistances and $V_a$ are the external voltages applied to the active coils.

In FGE, the profiles $p'$ and $TT'$ are expressed in terms of the linear combination of a set of basis functions $\mathcal{G}_P \cup \mathcal{G}_T$:
\begin{equation}
    p' = \sum_{g \in \mathcal{G}_p} a_{g} g', \quad
    TT' = \mu_0 \sum_{g \in \mathcal{G}_T} a_{g} g'.
\end{equation}
This enables the expression of the discretized current density using a matrix multiplication with the $a_g$ coefficient vector: $I_y = T_{yg}(\psi) a_g$, where the dependence of $T_{yg}$ on $\psi$ stems from evaluating the basis functions $g \in \mathcal{G}_P \cup \mathcal{G}_T$ on the normalized poloidal flux.

The basis coefficients are constrained using plasma parameters in a residual function dependent on the state of the plasma and the conductors
\begin{equation}
\label{eq:constraints}
    0 = \Lambda(a_g, \psi, I_e, \ldots).
\end{equation}

A variety of residuals are implemented to constrain the $p'$ and $TT'$ profiles based on known plasma parameters. Among them are constraints for the toroidal and poloidal beta $\beta_t, \beta_p$, the plasma thermal energy $W_k$, the safety factor on axis $q_A$, the internal inductance $\ell_i$ and the plasma current $I_p$. A comprehensive list of implemented constraint formulations is found in Appendix~\ref{sec:appendic_agcon}. Furthermore, different versions of current diffusion equations are available to constrain the profiles. These are discussed in more detail in the next Section~\ref{sec:cdes}. 

In summary, FGE simultaneously solves equations \eqref{eq:GS}, \eqref{eq:circuit}, \eqref{eq:constraints} employing an implicit time-stepping scheme.

\subsection{Current Diffusion Equations}
\label{sec:cdes}
By itself, the above set of equations does not include the effects of plasma resistivity on the equilibrium. It is possible to add Ohm's law dictating how resistive effects dampen and diffuse the plasma currents across flux surfaces. 
A canonical starting point for deriving current diffusion equations (CDEs) is the $\bfB$-parallel formulation of Ohm's law:
\begin{equation}
\label{eq:ohmslaw}
    \sigma_\| \bfE \cdot \bfB =  \bfj \cdot \bfB - \bfj_{ni}\cdot \bfB.
\end{equation}
Here, $j_{ni}$ denotes non-inductive contributions to the plasma current. Starting from Ohm's law, different assumptions and simplifications lead to additional equations which can constrain one or more degrees of freedom of the plasma profiles. In many transport codes like RAPTOR \citep{RAPTOR_code}, TORAX \citep{torax2024arxiv} or ASTRA \citep{astra}, equation~\eqref{eq:ohmslaw} is flux surface averaged to solve for the 1D profile evolution. However, as FGE is based on a cartesian grid, computing flux surface averages is relatively expensive and hence avoided when solving for the plasma equilibrium.

Instead, FGE implements a range of models for the plasma current evolution from a prescribed plasma conductivity using CDEs which avoid the use of flux surface averages. A distinction is made between 0D CDEs, constraining the evolution of the total plasma current and 1D CDEs, modeling the evolution of the full plasma current profile. We list the available options for CDEs implemented in FGE, starting with 0D cases:
\begin{itemize}
    \item \texttt{OhmTor} The plasma is modeled as a conductor where the loop voltage applied is given by the change in poloidal flux over the plasma domain:
    \begin{equation}
        0=\frac{1}{I_p} I_y^T \frac{\partial \psi}{\partial t} + R_p (I_p - I_{ni}).
    \end{equation}
    Here, $I_{ni}$ denotes the total toroidal non-inductive current which is supplied externally.
    \item \texttt{OhmTor-rigid} The plasma is modeled as a rigid conductor with a self-inductance and bulk resistance. The induced loop voltage is given by changing currents in external conductors:
    \begin{equation}
    \begin{aligned}
        0 =\ &L_p \dot{I}_p + M_{pe} \dot{I}_e + R_p (I_p - I_{ni}),\quad \text{with}\\
        &L_p \coloneqq \frac{1}{I_p^2} I_y^T M_{yy} I_y, \quad
        M_{pe} \coloneqq \frac{I_y^T}{I_p} M_{ye}.
    \end{aligned}
    \end{equation}
    \texttt{OhmTor-rigid} allows the option to compute $L_p$ dynamically or have it provided externally.
    \item \texttt{CDE-stationary-state} Assuming a stationary-state case where the change in flux is mostly uniform over the plasma domain admits a more accurate modeling of the plasma response based on internal profiles and a parallel conductivity profile:
    \begin{equation}
        0 = I_p - I_{ni} + \frac{\partial (\psi_A + \psi_B)}{\partial t} \frac{T_B}{4 \pi} \int_{0}^{\sPhi_B} \frac{\sigma_\|}{T^2} \dx{\sPhi}.
    \end{equation}
    First derived in \citep{carpaneses_thesis}, a brief derivation is also found in Appendix~\ref{sec:cde_derivs} for completeness.
\end{itemize}

To resolve internal dynamics of the current profile using 1D CDEs, FGE offers three approaches based on a weak formulation \citep{heumann_weak_formulation} to avoid expensive flux surface integration:
\begin{itemize}
    \item \texttt{CDE-1D} This CDE implements the weak formulation \citep{heumann_weak_formulation} using $g \in \mathcal{G}_T$ as test functions. In this way, all degrees of freedom of the $TT'$ profile are constrained by the residual:
    \begin{equation}
    \begin{aligned}
        0 = &\int_{A_\mathcal{P}} \frac{1}{R}  \frac{\partial T(\psi)}{\partial t} g(\psi) \dx{R} \dx{Z}\\
        + &\int_{A_\mathcal{P}} \frac{1}{R} \frac{\partial \psi}{\partial t} T(\psi) g'(\psi) \dx{R} \dx{Z}\\
        + &\int_{A_\mathcal{P}} \frac{2 \pi}{\sigma_\|} (j_\phi T(\psi) - j_{ni} R B_0) g'(\psi) \dx{R} \dx{Z}\\
        + &\int_{A_\mathcal{P}} \frac{\abs{\nabla \psi}^2 T'(\psi)}{\mu_0 \sigma_\| R}  g'(\psi) \dx{R} \dx{Z}.
    \end{aligned}
    \end{equation}
    Here, $A_P$ denots the plasma domain in the $(R,Z)$-plane and $j_{ni}$ is the non-inductive current density usually expressed through the flux surface average $\left\langle \bfj_{ni} \cdot \bfB\right\rangle / B_0$ which has to be supplied externally.
    \item \texttt{CDE-OhmTor-1D} A simpler, computationally less expensive formulation is obtained by either considering \texttt{CDE-1D} using the approximation $T\equiv const.$ or by integrating the toroidal projection of Ohm's law against profile test functions:
    \begin{equation}
    \begin{aligned}
        0 = &\int_{A_\mathcal{P}} \frac{1}{R} \frac{\partial \psi}{\partial t} g'(\psi) \dx{R} \dx{Z}\\
        + &\int_{A_\mathcal{P}} \frac{2 \pi}{\sigma_\phi} (j_\phi - j_{ni,\phi}) g'(\psi) \dx{R} \dx{Z}.
    \end{aligned}
    \end{equation}
    \item \texttt{static-CDE-1D} This CDE implements a static version of \texttt{CDE-1D} which acts like a constraint of the plasma current while allowing to solve for the relaxed stationary-state current density profile. For this formulation, all time derivatives in \texttt{CDE-1D} are omitted and $\partial_t \psi$ is assumed to be spatially uniform providing the loop voltage required to sustain the reference plasma current:
    \begin{equation}
    \begin{aligned}
        0 &= \mathcal{I}_{g_i}^{(1)} \mathcal{I}_{g_{i+1}}^{(2)} - \mathcal{I}_{g_i}^{(2)} \mathcal{I}_{g_{i+1}}^{(1)},\quad i=1,...,N_{T}-1\\
        0 &= \int_{A_\mathcal{P}} j_\phi \dx{R} \dx{Z} - I^{ref}_p
    \end{aligned}
    \end{equation}
    with $\{g_i\}_{i=1}^{N_T} \subseteq \mathcal{G}_T$ test functions and
    \begin{align}
        \mathcal{I}_{g_i}^{(1)} &= \int_{A_\mathcal{P}} \frac{2 \pi}{\sigma_\|} (j_\phi T(\psi) - j_{ni} R B_0) g_i'(\psi) \dx{R} \dx{Z}\notag \\
        &+ \int_{A_\mathcal{P}} \frac{\abs{\nabla \psi}^2 T'(\psi)}{\mu_0 \sigma_\| R}  g_i'(\psi) \dx{R} \dx{Z}, \\
        \mathcal{I}_{g_i}^{(2)} &= \int_{A_\mathcal{P}} \frac{1}{R} T(\psi) g_i'(\psi) \dx{R} \dx{Z}.
    \end{align}
    This CDE is also available when solving for the equilibrium at a single point in time, for example when initializing a FGE simulation.
\end{itemize}

\section{Implementation}
\label{sec:solving_the_problem}
The full system of equations is implemented as a root finding problem with a state representation containing information on both the plasma and external conductors. It is then solved using Newton iterations.

\subsection{State Variables and Residuals}
Commonly, the state of the axisymmetric plasma equilibrium is given by the poloidal flux $\psi$ as well as information on internal plasma profiles $p', TT'$ and external conductors $I_e$. Hence, a possible FGE state representation is given by
\begin{equation}
    x_\psi = (\psi,a_g,I_e).
\end{equation}
Alternatively, the state can be expressed in terms of plasma and external currents together with profile information
\begin{equation}
    x_{I_y} = (I_y, a_g, I_e).
\end{equation}
Since the plasma occupies only a limited region inside the vessel, the resulting sparsity of $x_{I_y}$ enables optimized computations in FGE routines. Furthermore, using $x_{I_y}$ allows to treat the boundary conditions of $\psi$ implicitly for a more concise residual formulation.
FGE offers the option to use either state representations and provides optimized algorithms for each.

The main residual for the root finding algorithm is subdivided into three parts: One for the plasma representing the Grad-Shafranov equation, one for the circuit equation of external conductors and one for the $a_g$ constraints:
\begin{equation}
\label{eq:residual}
F(x) = \begin{pmatrix}
    F_{p}(x)\\
    M_{ey} \dot{I}_y + M_{ee} \dot{I}_e + R_e I_e - V_e\\
    \Lambda(a_g,\ldots)
\end{pmatrix}\!.
\end{equation}
Multiple options for the plasma residual formulation are offered, depending on the chosen state representation:
\begin{align}
    F_{p1}(x_{I_y}) &= I_y - I_y(\psi(I_y, I_e), a_g), \\ F_{p2}(x_\psi) &= \psi - \psi(I_y(\psi,a_g), I_e), \\ F_{p3}(x_\psi) &= \begin{cases}
        \dlst \psi + 2 \pi R \mu_0 I_y(\psi, a_g),\\
        \psi_b - M_{by} I_y - M_{be} I_e.
    \end{cases}
\end{align}
Here, $\psi(I_y, I_e)$ refers to the operation of inverting the $\dlst$-operator to compute the poloidal flux from the toroidal current density. It uses Dirichlet boundary conditions given by the influence of plasma and external currents on the boundary. We note that this inversion is rapidly and efficiently implemented using a tridiagonal matrix inversion subroutine. Conversely, $I_y(\psi, a_g)$ refers to the operation of finding the plasma boundary by finding limiting- and X-points and recomputing $I_y=T_{yg}(\psi) a_g$.

A notable part of the residual concerns the identification of the plasma domain via a search for X-points and limiting points. X-point quantities are computed based on a 6-point stencil method outlined in \citep{liuqe_paper} where X-point flux and position are found using a local quadratic fit. Depending on user settings, this fit is improved based on a spline interpolation of the flux map. Limiting points are treated similarly based on local extremum approximations with various available methods ranging from bilinear stencils to spline fits.

To ensure numerical stability, FGE normalizes both the residuals and state variables to unit magnitude based on tokamak-specific knowledge and scalings.

\subsection{Newton Method}
A multitude of Newton-type algorithms for finding roots of the above residuals are implemented, all based on the general update rule
\begin{equation}
    J_F \delta x = -F(x).
\end{equation}
The inversion of the jacobian $J_F$ is performed either using a Jacobian-Free Newton-Krylov (JFNK) algorithm, where the action of the jacobian is approximated using finite differencing, or by directly using the full jacobian matrix. The latter approach is made efficient by the availability of analytical derivatives for all residuals in the code. Jacobians of limiting- and X-point quantities are obtained by analytic differentiation of the local fits used to compute them. We refer to \citep[Appendix C]{carpaneses_thesis} for a more detailed description.

Time derivatives in residual~\eqref{eq:residual} arising from circuit equations and CDEs are usually handled by finite differences with respect to the last timestep. For some applications however, dependencies of the residual on state variables and their time derivatives need to be separated. For these cases, analytical jacobians are available for either.

Several GMRES \citep{gmres_orig} implementations are offered to handle the efficient inversion of $J_F$. These include the MATLAB-native GMRES solver as well as custom implementations that handle the inversion of the Householder matrix via the MATLAB backslash operator, Givens rotations or Ayachour's method \citep{AYACHOUR2003269}. If the full jacobian is known, a LU decomposition or the MATLAB backslash operator are also available for the inversion. Moreover, when using GMRES implementations in conjunction with the analytical jacobian, FGE allows to compute analytic jacobian-vector products implicitly without the need to assemble the full matrix resulting in notable performance improvements for many cases.

Preconditioning in FGE can be performed by computing the full inverse of the jacobian, using an incomplete LU factorization, or by supplying custom preconditioning matrices externally. Both left and right preconditioning is supported by the custom GMRES implementations. Users can also choose when the preconditioning matrices are recomputed, either once at the start of the simulation, before each time step, or before each Newton iteration. The required level of preconditioning depends on the chosen residual formulation. Specifically, residuals $F_{p1}$ and $F_{p2}$, which already incorporate the inversion of the $\dlst$-operator, yield jacobians that are generally well-conditioned and diagonally dominant, requiring only minimal preconditioning for convergence. In contrast, the jacobian of residual $F_{p3}$ benefits significantly from preconditioning for the efficient inversion through GMRES iterations.

Lastly, the solvers implemented in FGE are optimized to exploit known sparsity patterns in the problem. As $I_y$ is constrained to be zero outside of the plasma domain, the jacobian of the residual $F_{p1}$ usually contains many rows with only one non-zero element on the diagonal which are trivially eliminated, allowing for a significant reduction of the system size through masking.
However, because $F_{p1}$ incorporates the inverse of the $\dlst$-operator, the remaining jacobian is generally dense. 
While not allowing for the same system reduction, the $F_{p3}$ formulation features a jacobian that is inherently sparse as the local $\dlst$-operator is applied to compute the residual but is never inverted. For this case, all analytical jacobian computations, preconditioning, and GMRES algorithms are equipped to fully leverage the sparsity to speed up the solver.

\section{Linearization of the Evolution\\ Operator}
\label{sec:linearization}
For control applications, it is often essential to express a linearized dynamics in the form of a state-space system to be used for controller design, gain tuning and system analysis. FGE offers the capability to linearize the evolution operator around a converged equilibrium to yield a plasma response model in a state-space representation which can be viewed as a generalization of the methods outlined in \citep{Walker2006}. The resulting state-space system describes how perturbations from the initial equilibrium evolve in time and in response to control inputs. It serves to analyze the effects of control actions and allows estimation of vertical instability growth rates by computing unstable eigenmodes. Additionally, FGE provides the possibility to carry out full plasma simulations with controllers in the loop based on this linearized representation allowing for rapid prototyping and gain tuning. As will be seen in Section~\ref{sec:fgel_comp_subsec}, as long as the plasma stays close to the initial equilibrium, the linearized system reproduces the full non-linear dynamics very well at a fraction of the computational cost.

To obtain a light-weight linearized system representation,
we exploit the fact that not all parts of the residual depend on time derivatives of the state. Here, we explicitly include the dependence of the residual from equation~\eqref{eq:residual} on the state representation $x$, its time derivative $\dot{x}$ and the control inputs $u$ containing applied coil voltages, non-inductive current drive and $a_g$ constraint references. $F(x,\dot{x},u)$ is split into static residuals $F_S(x, u)$ and dynamic residuals $F_D(x, \dot{x}, u)$ based on their dependence on $\dot{x}$.
Likewise, the state vector $x$ is divided into static $x_S$ and dynamic $x_D$ components. This subdivision is not inherently predetermined but must fulfill the condition that the resulting jacobian $\partial F_S / \partial x_S$ has full rank. A canonical choice for $x_S$ and $x_D$ is to group quantities involved in the force balance equation ($I_y$ or $\psi$) as static states and quantities related to circuit equations ($I_e$) as dynamic states with states containing the $a_g$ coefficients being assigned to either group depending on the usage of plasma constraints and CDEs.

\begin{figure}[!b]
    \centering
    \includegraphics[width=\linewidth]{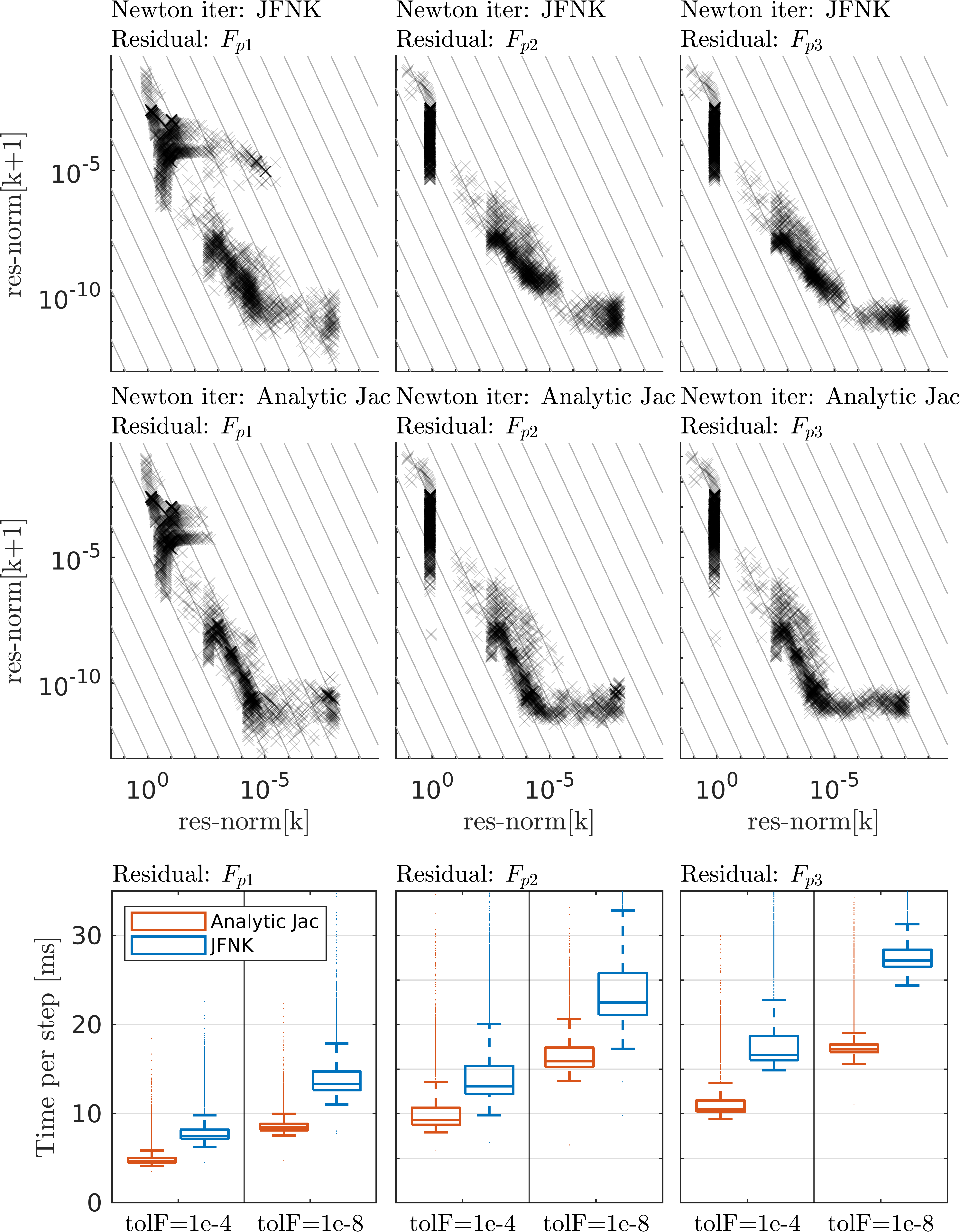}\\
    \caption{Comparison of convergence rate and computational time between the three residuals for the newton solver using the analytical jacobian and the JFNK solver. Data is taken from a FGE simulation of TCV \#61400 from 0.1s to 0.8s done at 10kHz.
    The top plot shows residual norms of the k-th iteration vs the k+1-th iteration with tolF=1e-8. Only the first 1000 time steps are shown in the scatterplot to maintain visual clarity. The background lines indicate second order convergence. The bottom shows box-plots of the computation time spent per time step for two different solver tolerances using an Intel i9-12900K CPU. The box-plots display the 25th to 75th percentile as the box with the median line inside. The whiskers extend to 1.5 times the inter-percentile range.}
    \label{fig:convergence}
\end{figure}

\begin{figure}[!b]
    \centering
    \includegraphics[width=\linewidth]{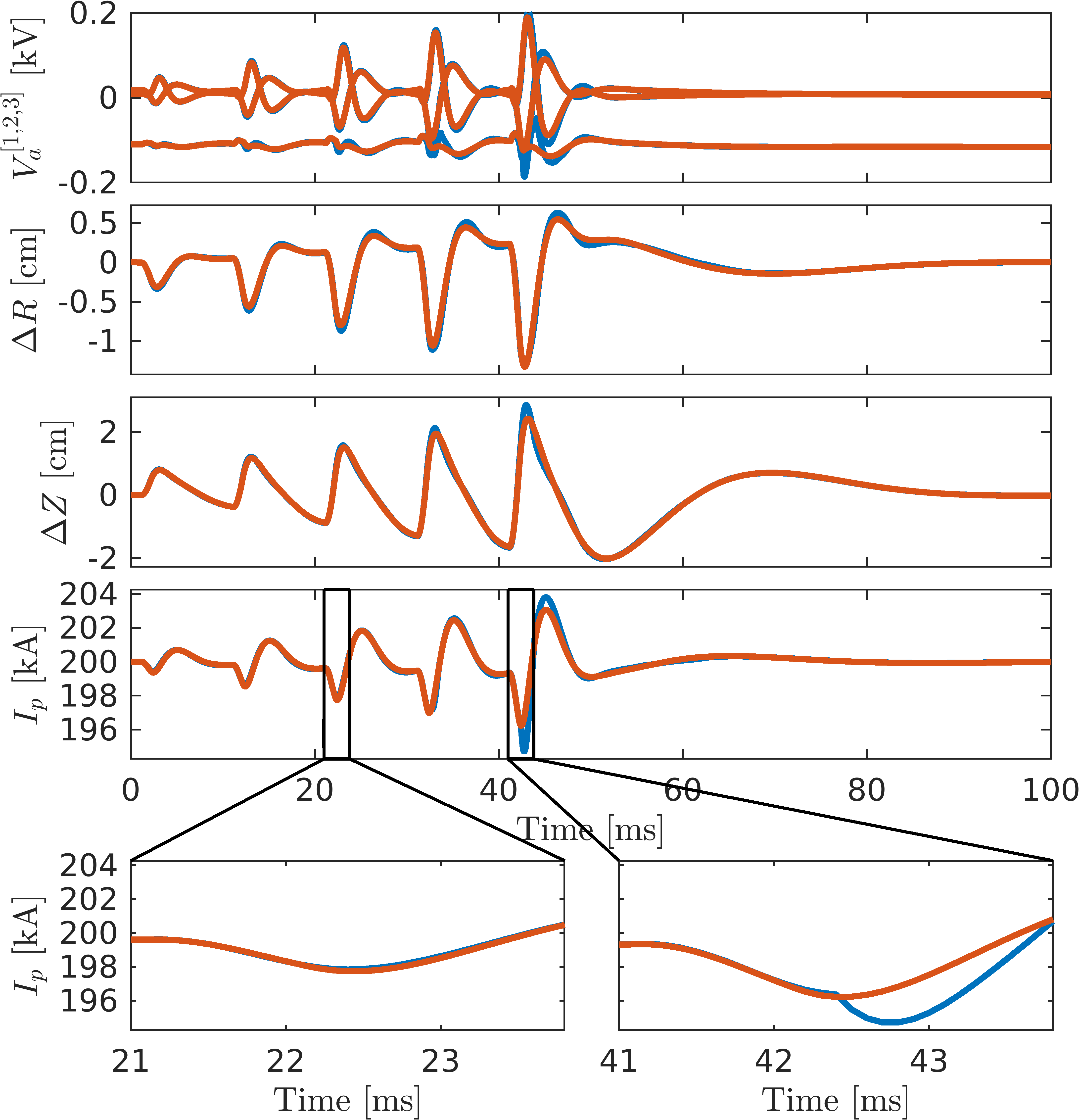}\\
    \vspace{0.2cm}
    \includegraphics[scale=0.59]{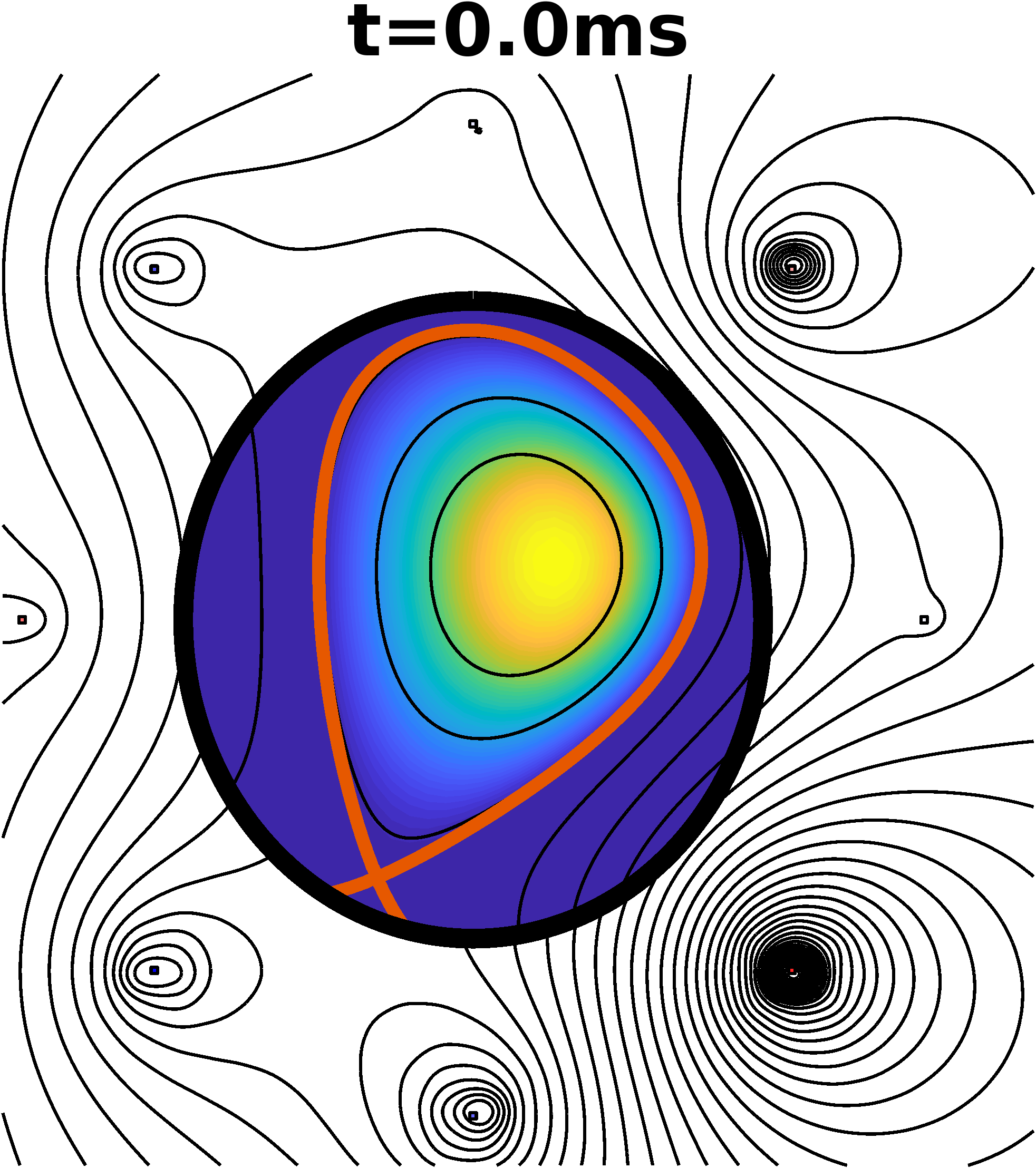}
    \hspace{0.4cm}
    \includegraphics[scale=0.59]{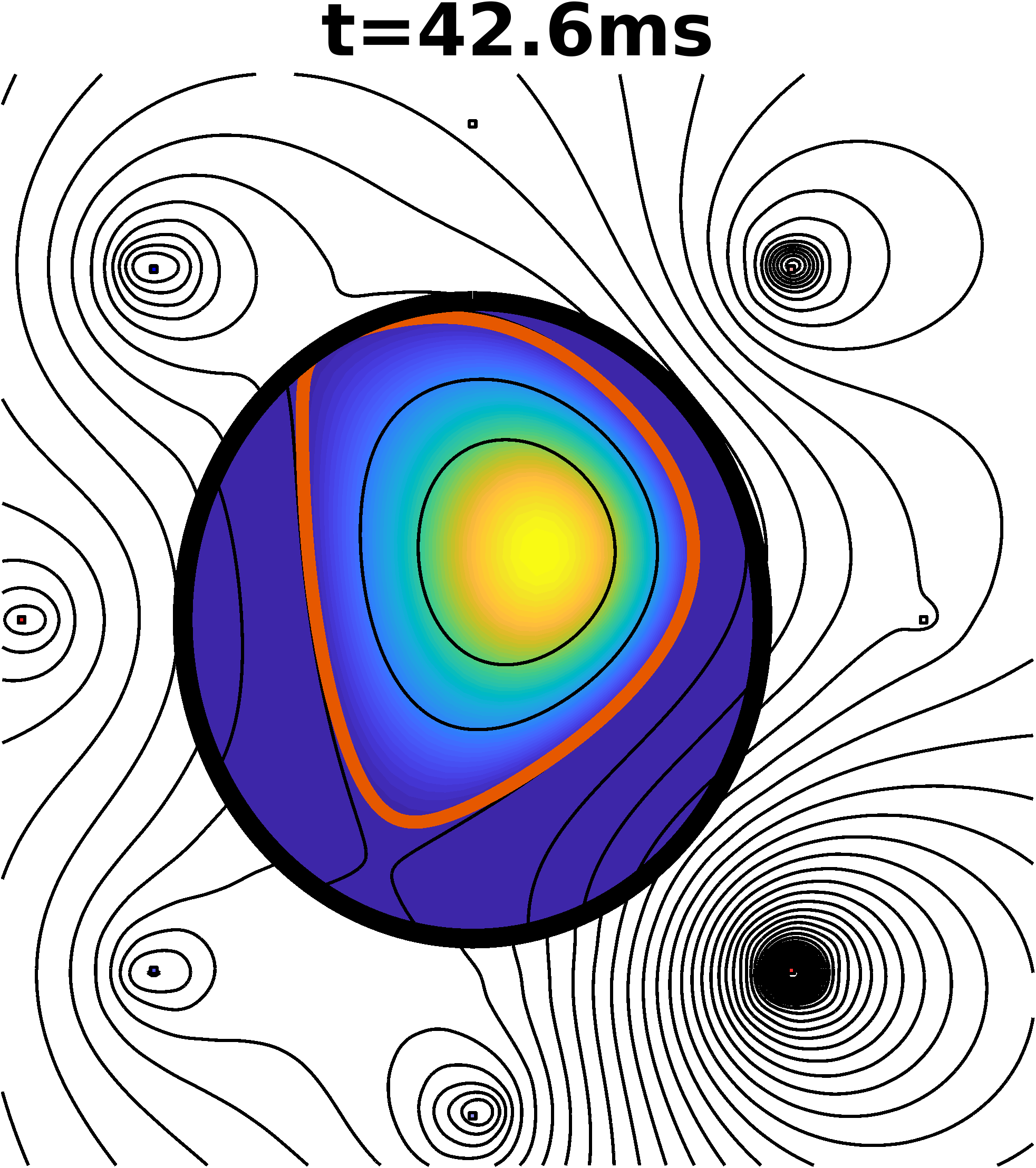}
    \caption{Comparison between full nonlinear simulation (blue) and state-space model (orange) for Anamak \#2 (see Appendix~\ref{sec:anamak}). Traces for the coil voltages applied to the first three coils are shown at the top. Below, radial and vertical displacement and the plasma current are shown. At the bottom, two zooms of the plasma current trace are provided. Finally, two equilibria from the nonlinear simulation at 0$ms$ and 43$ms$ are shown where contours of the poloidal flux are plotted in black, the LCFS is highlighted in red and the toroidal current distribution is shown in color.}
    \label{fig:enter-label}
\end{figure}

Given a static equilibrium $x_0, u_0$ with $0 = F(x_0, \dot{x}_0, u_0)$ and $\dot{x}_0=0$, the residual is expanded to arrive at the linearized representation
\begin{align}
\label{eq:linearized_root_equation}
    0 &= \frac{\partial F_S}{\partial x_S} \deltax_S + \frac{\partial F_S}{\partial x_D} \deltax_D + \frac{\partial F_S}{\partial u} \deltau,\\
    0 &= \frac{\partial F_D}{\partial \dot{x}_S} \dot{\deltax}_S + \frac{\partial F_D}{\partial \dot{x}_D} \dot{\deltax}_D\notag \\
    &+ \frac{\partial F_D}{\partial x_S} \deltax_S + \frac{\partial F_D}{\partial x_D} \deltax_D + \frac{\partial F_D}{\partial u} \deltau.
\end{align}
Here, the perturbations $\deltax, \deltau$ are relative to the reference equilibrium and all jacobians are computed at $(x_0, u_0)$.
Equation~\eqref{eq:linearized_root_equation} now admits an expression of the static states in terms of the dynamic ones:
\begin{align}
    \deltax_S &= \frac{\partial x_S}{\partial x_D} \deltax_D + \frac{\partial x_S}{\partial u} \deltau, \quad \text{with}\\
    \frac{\partial x_S}{\partial x_D} &\coloneqq -\left(\frac{\partial F_S}{\partial x_S}\right)^{-1} \frac{\partial F_S}{\partial x_D}, \\
    \frac{\partial x_S}{\partial u} &\coloneqq -\left(\frac{\partial F_S}{\partial x_S}\right)^{-1} \frac{\partial F_S}{\partial u}.
\end{align}
Defining
\begin{align}
\begin{aligned}
    S &\coloneqq \frac{\partial F_D}{\partial \dot{x}_D} + \frac{\partial F_D}{\partial \dot{x}_S} \frac{\partial x_S}{\partial x_D},\\
    U &\coloneqq -\frac{\partial F_D}{\partial u} - \frac{\partial F_D}{\partial x_S} \frac{\partial x_S}{\partial u},
\end{aligned}
\quad
\begin{aligned}
    K &\coloneqq -\frac{\partial F_D}{\partial x_D} - \frac{\partial F_D}{\partial x_S} \frac{\partial x_S}{\partial x_D},\\
    V &\coloneqq -\frac{\partial F_D}{\partial \dot{x}_S} \frac{\partial x_S}{\partial u},
\end{aligned}
\end{align}
the dynamic equation is expressed as
\begin{align}
    0 = S \dot{\deltax}_D - K \deltax_D - U \deltau - V \dot{\deltau},
\end{align}
Hence, the linearized evolution is governed by a state-space system ($A,B,C,D$) with
\begin{equation}
\begin{aligned}
    A &\coloneqq S^{-1} K, &B &\coloneqq [S^{-1}U, S^{-1}V], \\
    C &\coloneqq H \begin{bmatrix}
        \frac{\partial x_S}{\partial x_D}\\
        \mathbbold{1}_D
    \end{bmatrix}, &D &\coloneqq 0.
\end{aligned}
\end{equation}
Here, $H$ denotes the matrix extracting the measurements $y$ from the full state $x$ such that the dynamics are given by
\begin{equation}
\label{eq:state_space_sys}
\begin{aligned}
    \dot{\deltax}_D &= A \deltax_D + B {\delta}\hat{u}\\
    y &= C \deltax_D + D {\delta}\hat{u} + y_0,
\end{aligned}
\end{equation}
where $y_0$ are the measurements obtained from the initial equilibrium $x_0$ and ${\delta}\hat{u}$ denotes the vector $[\deltau, \dot{\deltau}]$. Expressing the linearized system only in terms of the dynamic states allows for a drastic reduction in system size resulting in the number of states being determined mainly by the number of degrees of freedom used to describe external currents, independently of the grid used.

\begin{figure*}[!ht]
    \centering
    \def\plotwidthkc{0.37}
    \begin{tabular}{c|c}
        FGE & KINX\\
        \vspace{-0.2cm} & \vspace{-0.2cm}\\
        \includegraphics[scale=\plotwidthkc]{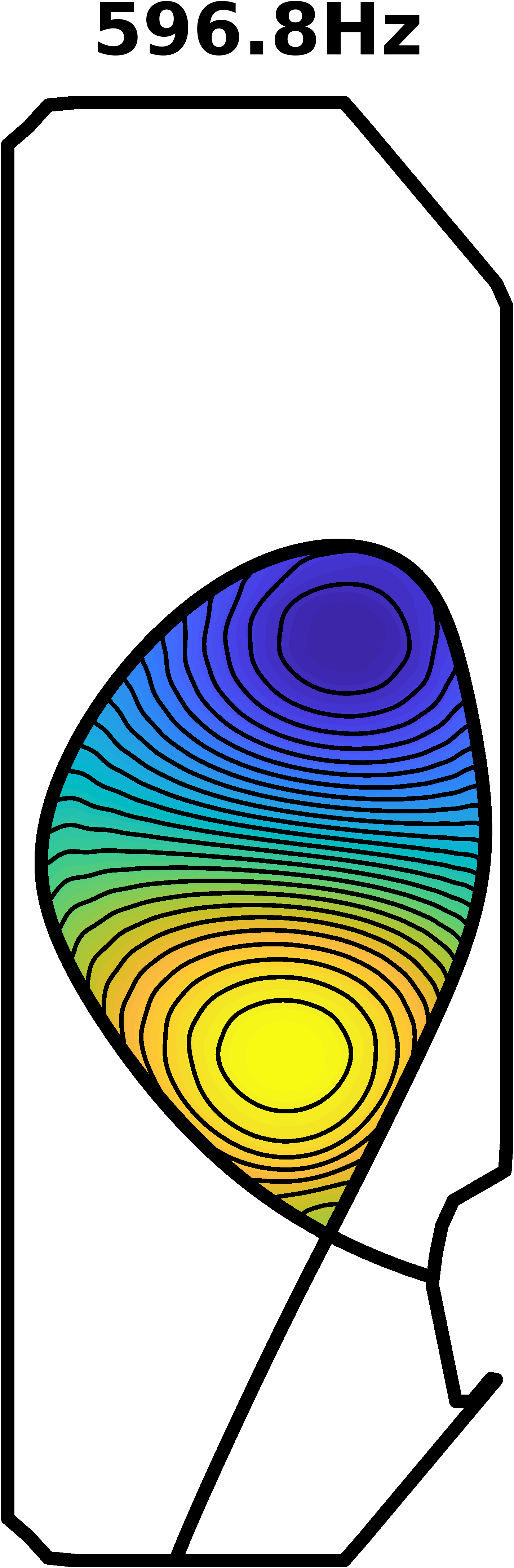}
        \includegraphics[scale=\plotwidthkc]{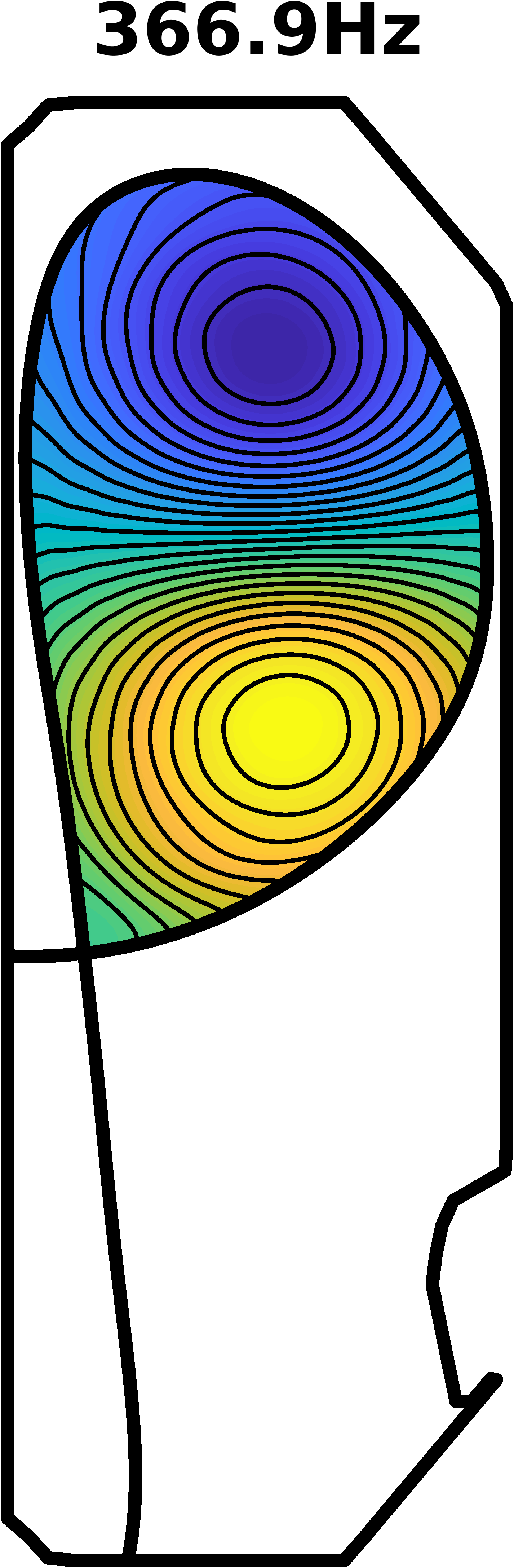}
        \includegraphics[scale=\plotwidthkc]{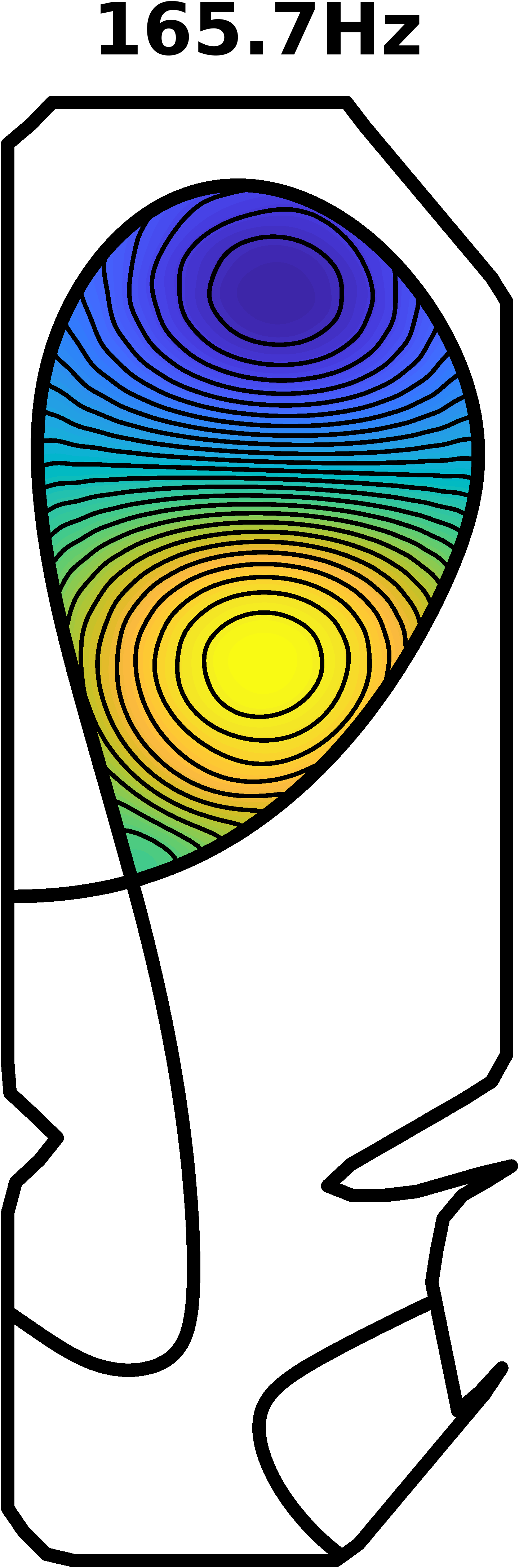}
        \hspace{0.2cm}
        \includegraphics[scale=\plotwidthkc]{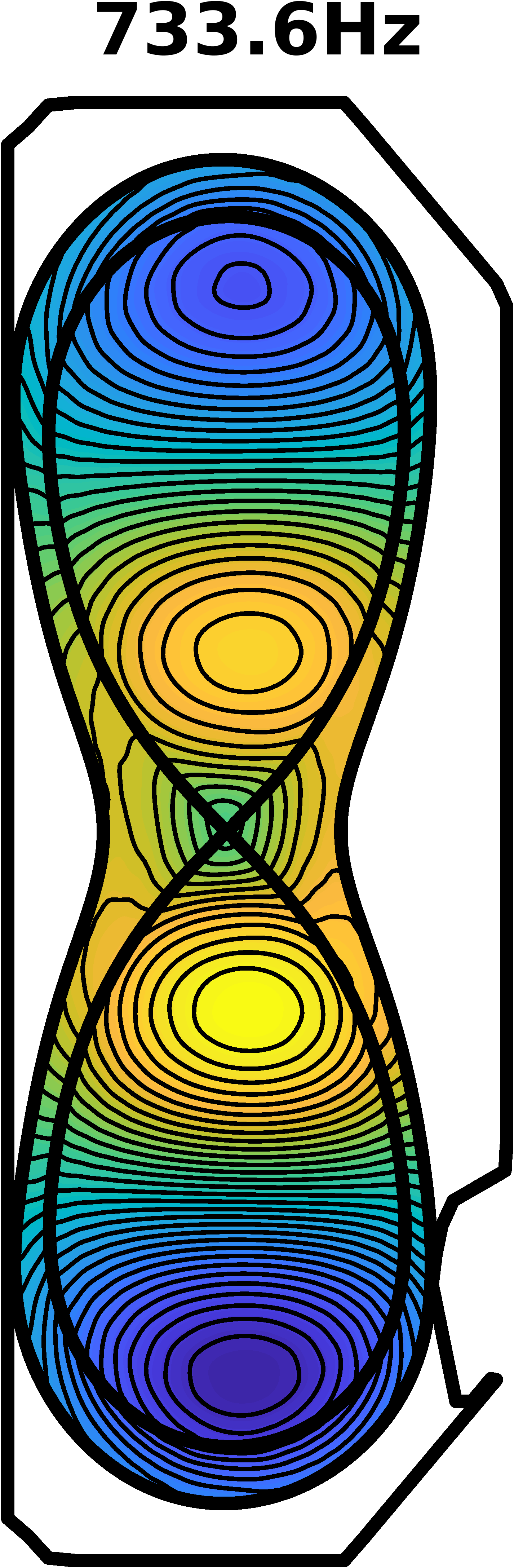}
        \includegraphics[scale=\plotwidthkc]{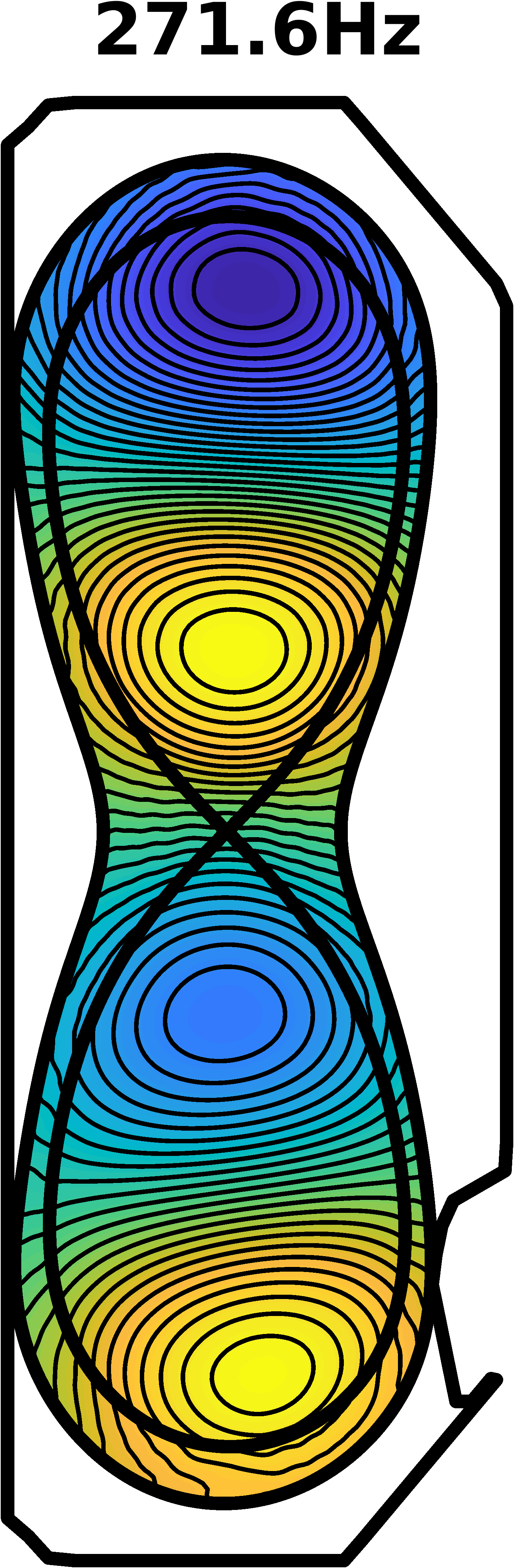}
        & 
        \includegraphics[scale=\plotwidthkc]{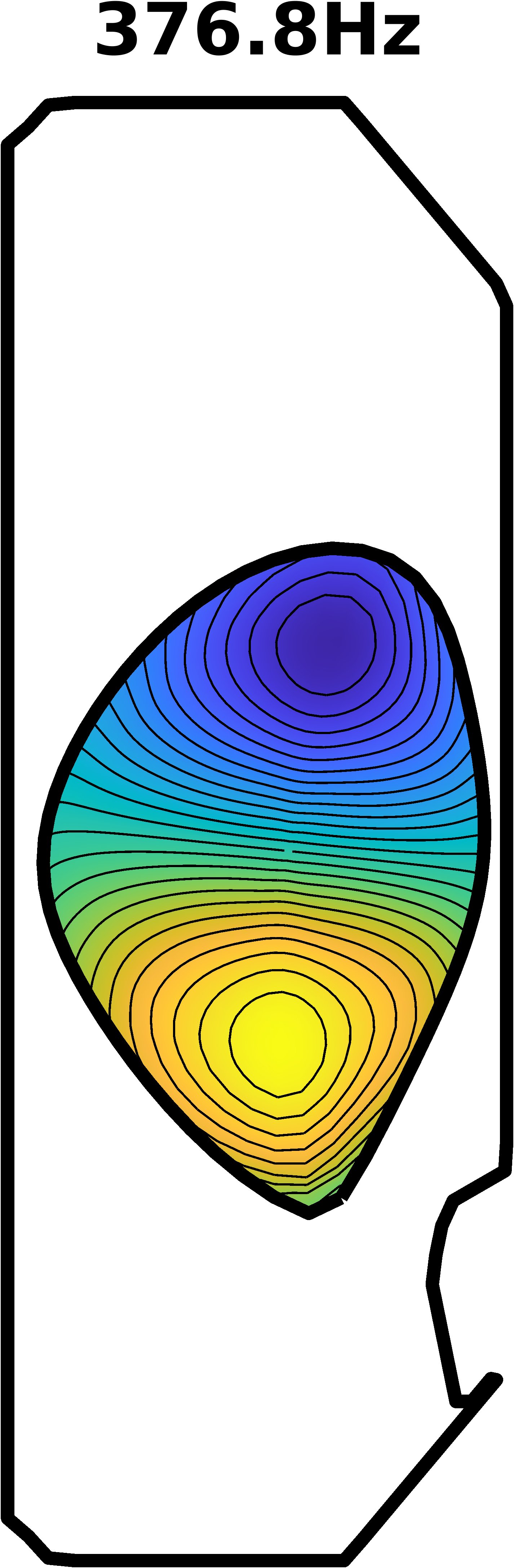}
        \includegraphics[scale=\plotwidthkc]{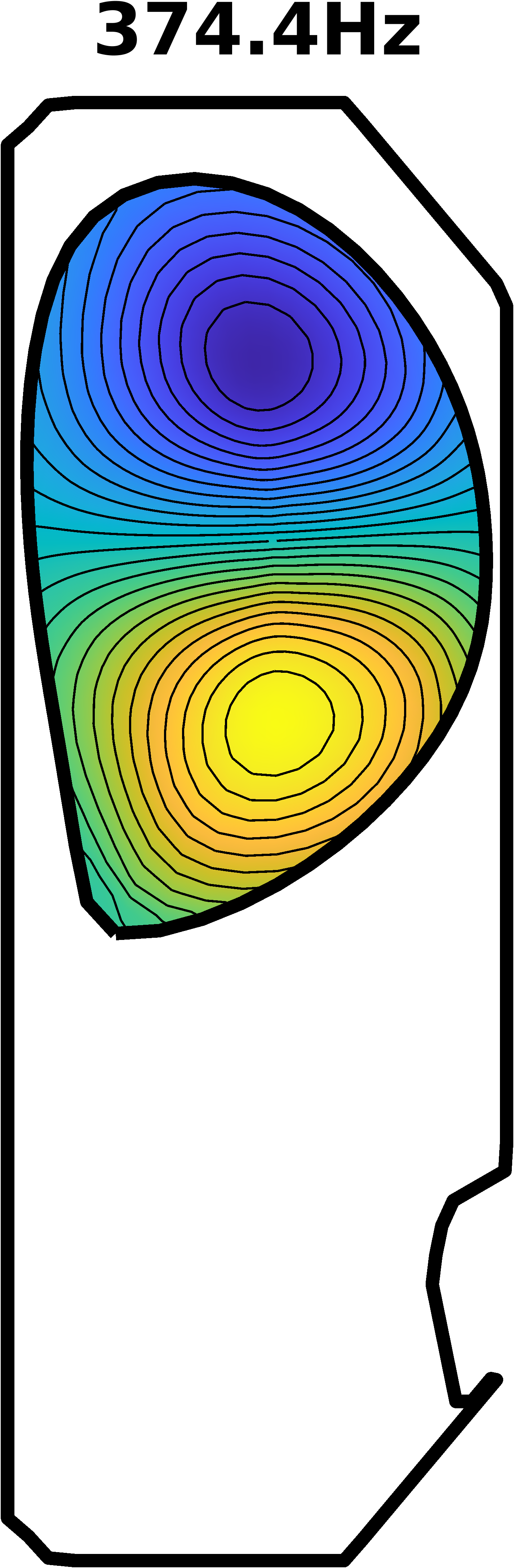}
        \includegraphics[scale=\plotwidthkc]{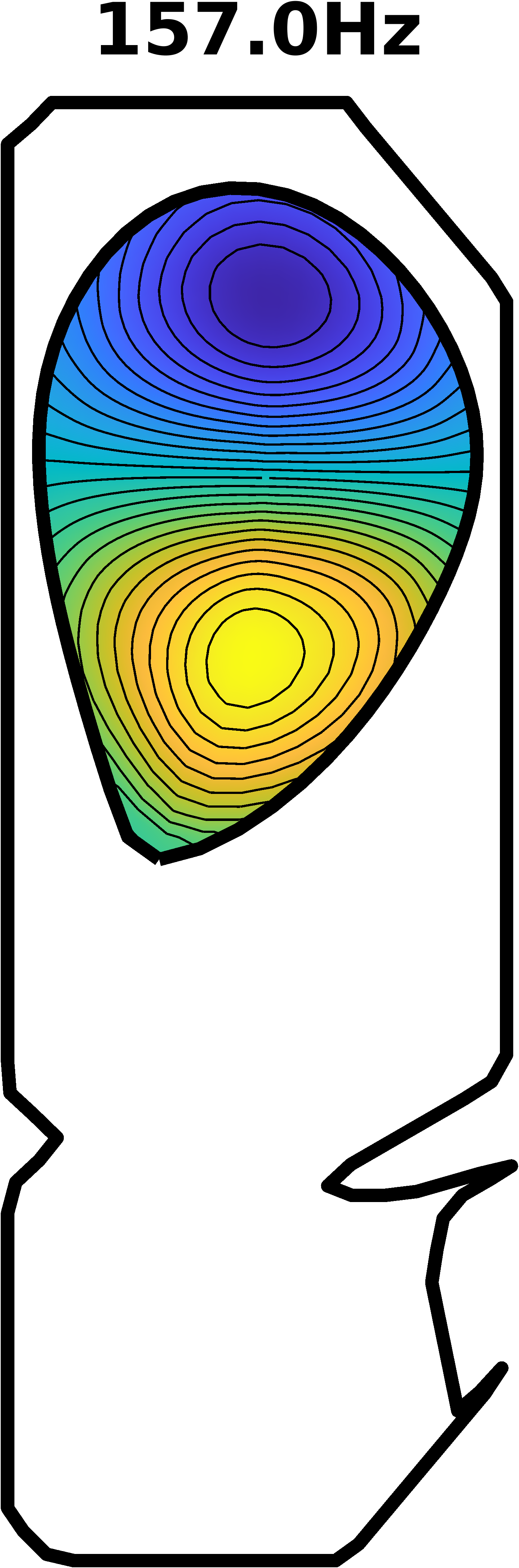}
        \hspace{0.2cm}
        \includegraphics[scale=\plotwidthkc]{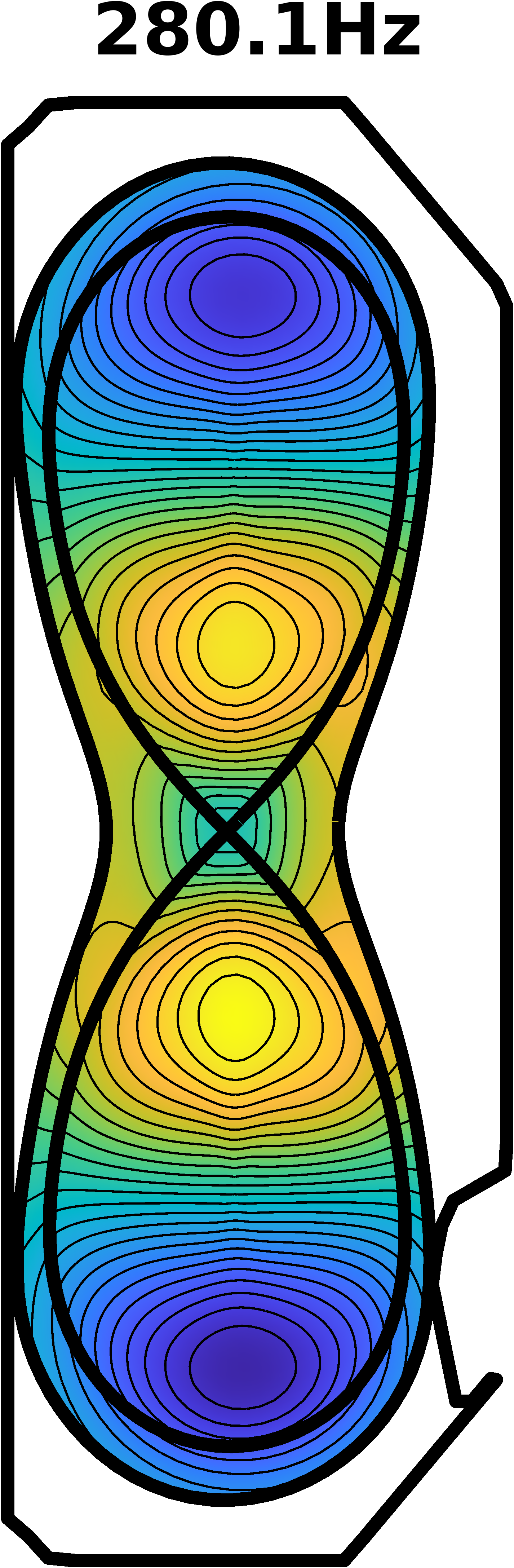}
        \includegraphics[scale=\plotwidthkc]{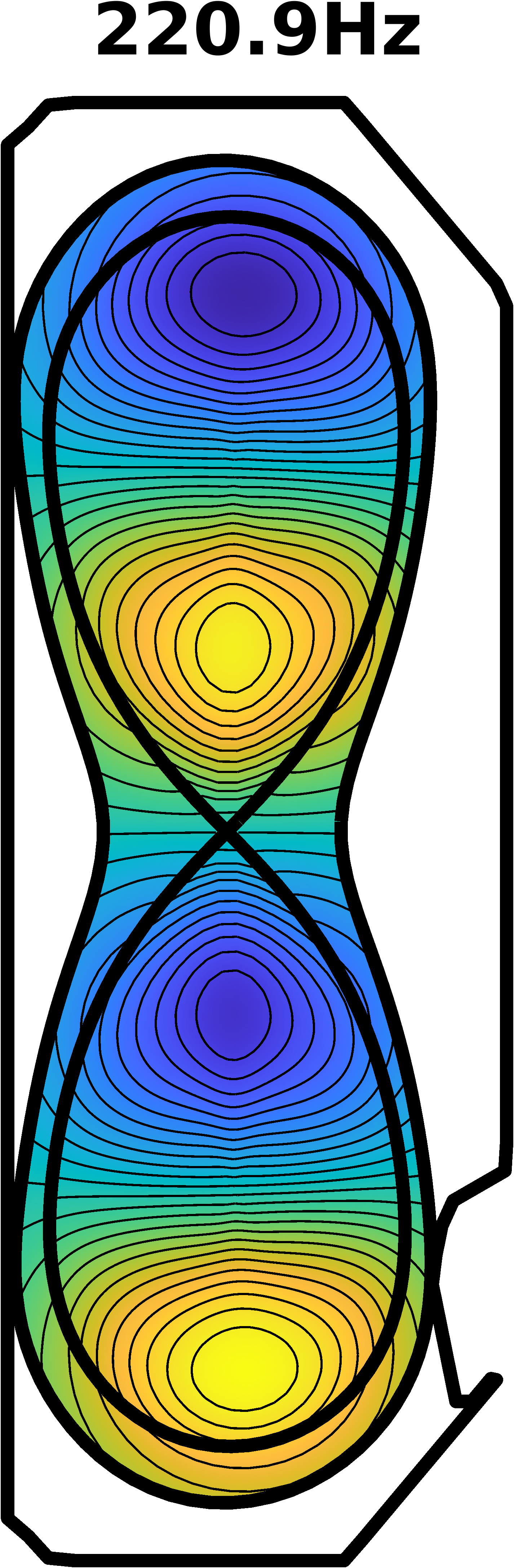}
    \end{tabular}
    \caption{Comparison of FGE (left) and KINX (right) for vertical growth rates of different plasmas. From left to right: TCV shots \#67084, \#77017, \#84062 at 0.9s plus a fictitious doublet plasma. The poloidal flux perturbation of the largest Eigenmodes is shown inside the plasma domain.}
    \label{fig:kinx_comp}
\end{figure*}

\begin{figure}[!hb]
    \centering
    \hspace{0.5cm}
    \includegraphics[width=0.9\linewidth]{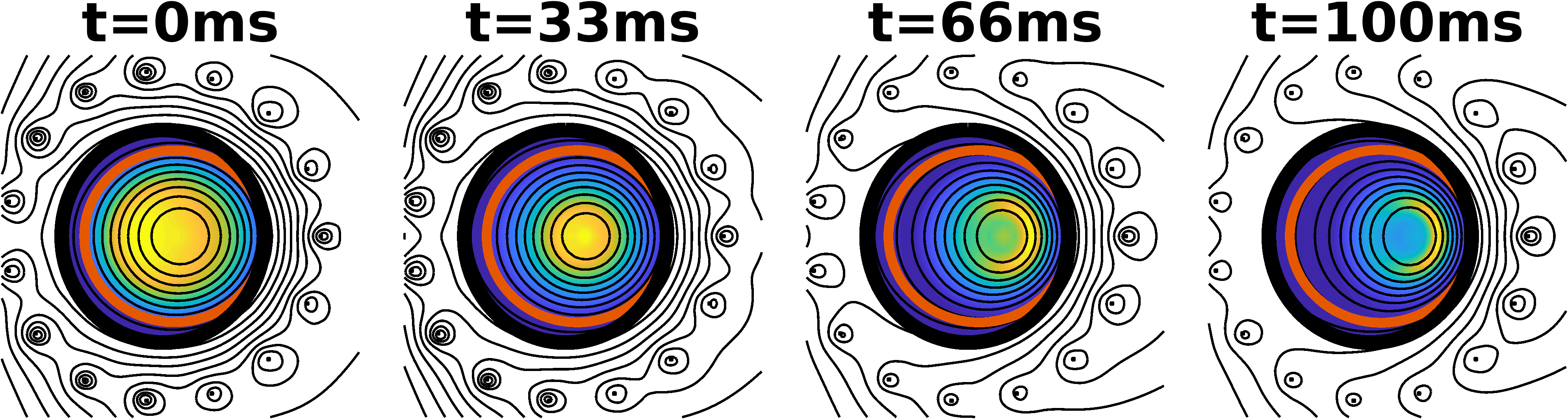}\\
    \vspace{0.4cm}
    \includegraphics[width=\linewidth]{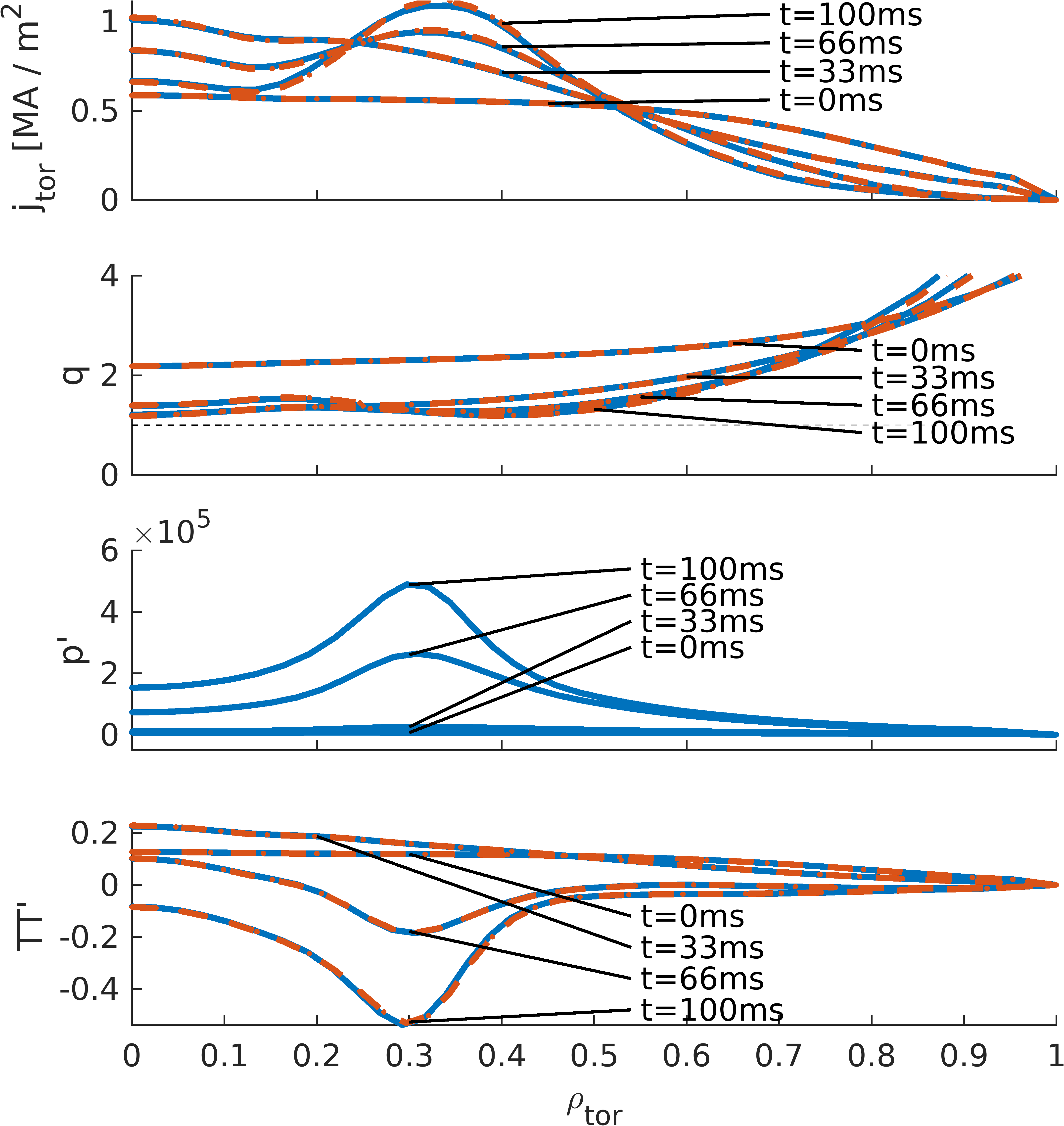}
    \caption{Comparison between FGE using the \texttt{CDE-1D} and the FBT-RAPTOR coupling for Anamak \#1 (see Appendix~\ref{sec:anamak}). The simulation is run for 100$ms$ and off-axis ECCD is switched on after 30$ms$. The top equilibria show contours of the poloidal flux with the toroidal current density drawn in color showing the transition from a flat current profile to an off-axis peaked current distribution. Below, profiles for current density, $q$, $p'$ and $TT'$ are shown for FGE (blue) and FBT-RAPTOR (orange) for different times. The $p'$ profiles are only plotted in blue as they constrained to be equal between FGE and FBT-RAPTOR.}
    \label{fig:raptor_comp_dynamic}
\end{figure}

\begin{figure}[!hb]
    \centering
    \includegraphics[width=\linewidth, trim={0 0 0 0.32cm},clip]{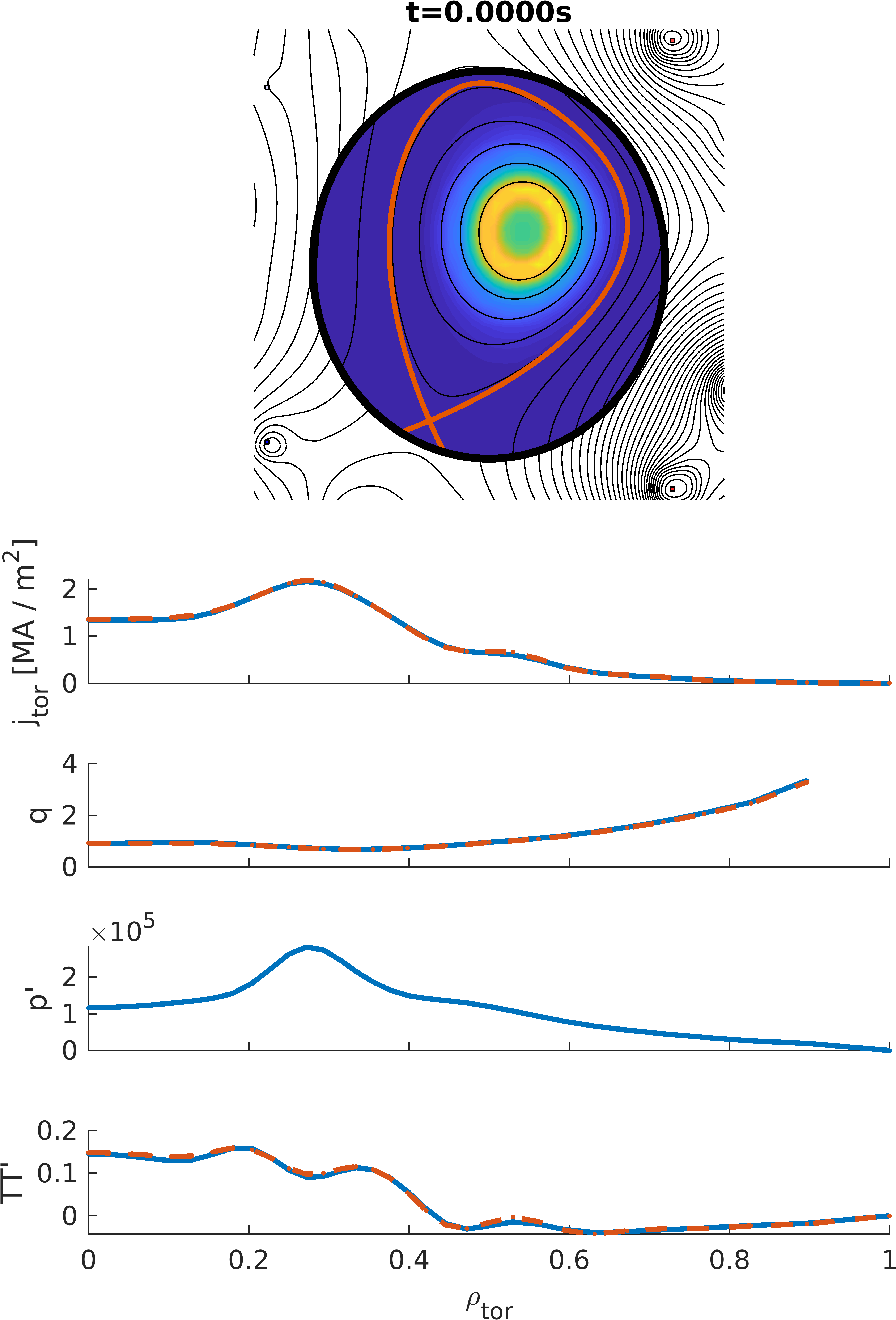}
    \caption{Comparison between the FGE initialization using the stationary-state 1D CDE (blue) and the raptor stationary-state solver (orange) for Anamak \#2 (see Appendix~\ref{sec:anamak}) with off-axis ECCD. The equilibrium from the FGE initialization is shown at the top with poloidal flux contours plotted in black, the LCFS highlighted in red and the toroidal current density drawn in color. Below, plasma profiles are compared where $p'$ is only plotted in blue as it is constrained to be equal between FGE and FBT-RAPTOR.}
    \label{fig:raptor_comp_static}
\end{figure}

\begin{figure}[!b]
    \centering
    \hspace{0.4cm}
    \includegraphics[width=0.93\linewidth]{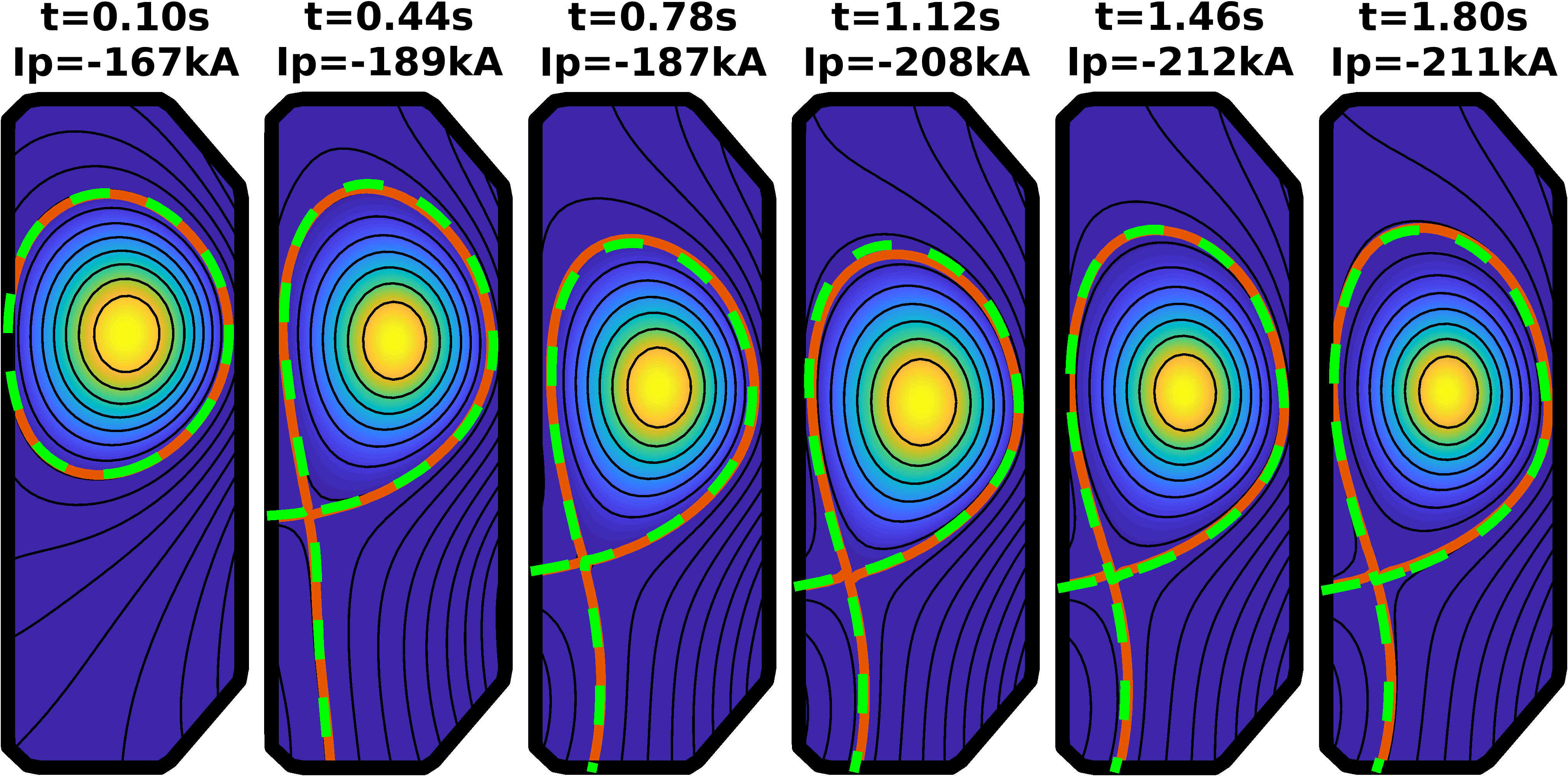}
    \vspace*{0.2cm}\\
    \includegraphics[width=\linewidth]{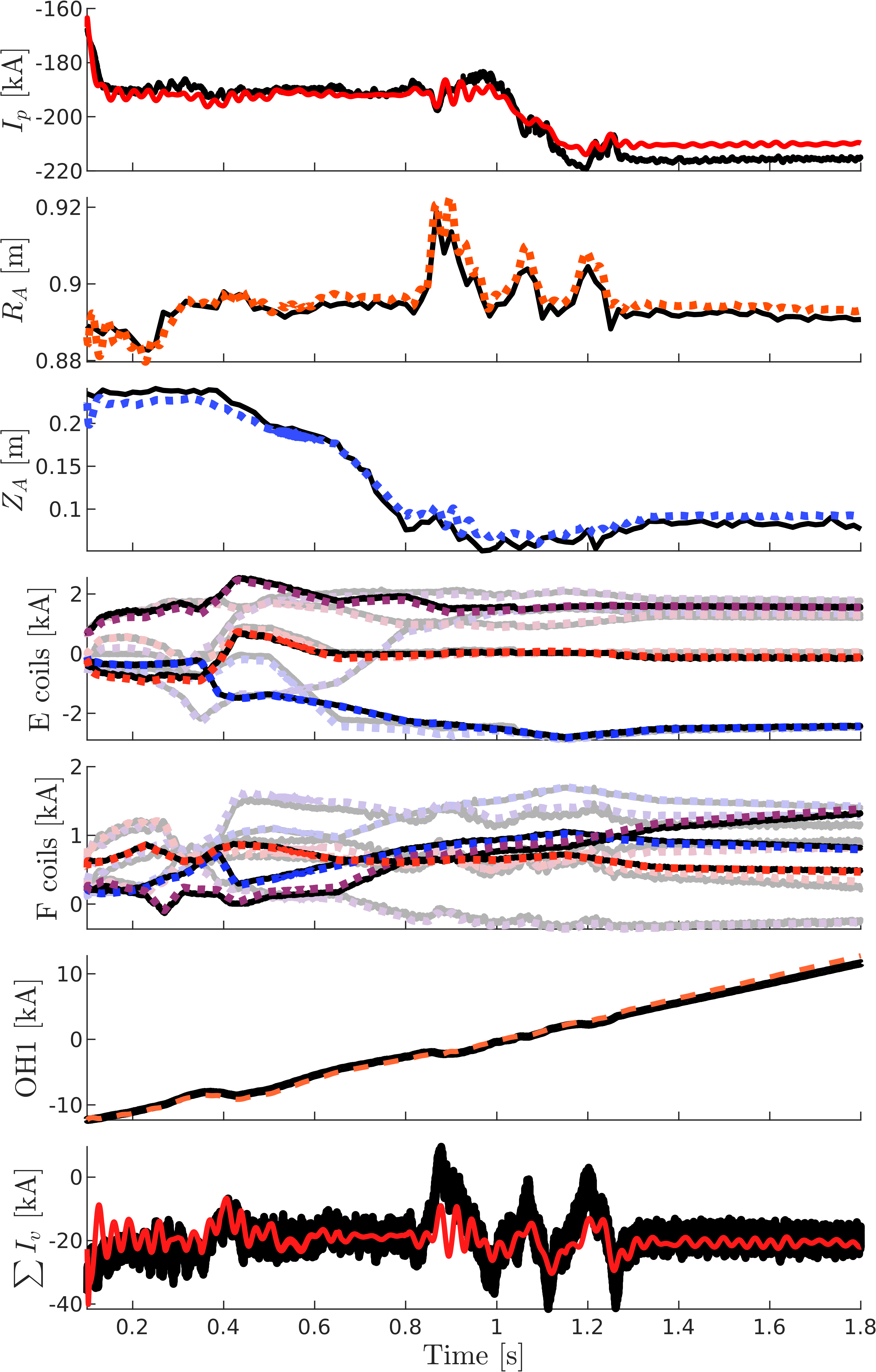}
    \caption{Results from the FGE resimulation of TCV \#61400. The top plot shows selected equilibria from the simulation. The LCFS from the LIUQE reconstruction is overlayed in green. The time traces below show a comparison between simulated quantities (black) and measurements (colored). Vertical lines highlight the times of the plotted equilibria. Measurements of axis position $R_A$, $Z_A$ and vessel filament currents $I_v$ are taken from the LIUQE reconstruction.}
    \label{fig:resim}
\end{figure}

\section{Numerical Studies}
\label{sec:results}
This section presents results from numerical studies demonstrating the performance of FGE. Apart from presenting results on the fictitious Anamak tokamak (see Appendix~\ref{sec:anamak}), we use TCV \#61400 for our numerical studies. TCV \#61400 was chosen as a benchmark case as it contains shape changes throughout the shot, EC and NBI heating and an unusual transition from a limited to a diverted configuration as the active X-point is formed very close to the limiter. We first examine solver convergence, verifying expected convergence rates and reporting computational times. Next, we provide a comparison between the full nonlinear evolution with its linearized counterpart. We benchmark vertical growth rates against the KINX code \citep{kinx_code_paper} and evaluate the 1D CDE implementations against a RAPTOR baseline. Finally, we present a measurement based resimulation of TCV \#61400 using FGE.

\subsection{Convergence Studies}
\label{sec:convergence_studies}
To verify second order convergence for the different solver methods and residuals, a simulation of TCV \#61400 is performed using various solver settings with a time step of 100$\mu s$. Specifically, we compare the use of all three presented residuals as well as comparing the JFNK method with the Newton method based on the analytical jacobian. In all cases, we use the GMRES implementation from \citep{AYACHOUR2003269} and enable matrix-free jacobian-vector product computations. For residuals $F_{p1}$ and $F_{p2}$, we employ left preconditioning using the inverse jacobian from the first time step, while for $F_{p3}$, preconditioning is done using the incomplete LU factorization, recomputed at every time step. For the plasma profiles, the \texttt{OhmTor} CDE is used with $q_A$ and $\beta_p$ constrained to reference values from the shot preparation. A simple position and $I_p$ controller is employed to keep the plasma stable 

Figure~\ref{fig:convergence} shows the results from this comparison. All methods generally exhibit second-order convergence, though the JFNK approach shows slightly diminishing convergence rates as residuals become very small. The achievable minimum residual norm is consistently bounded by floating point errors, with the specific floor varying across residual formulations. On average, the Newton solver requires approximately 2.1 iterations per time step for a tolerance $\text{tolF}{=}10^{-8}$ and 1.2 iterations for $\text{tolF}{=}10^{-4}$, where $\text{tolF}$ is the tolerance for the 2-norm of the normalized residual used in the Newton solver.
Additionally, Figure~\ref{fig:convergence} shows the time spent per time step for the simulation of TCV \#61400 with different residual formulations and solver methods. The simulation is carried out using the standard TCV grid ($28 {\times}65$) on a consumer-grade desktop computer using an Intel i9-12900K CPU. For well chosen settings, the average computational time spent per time step is less than 10$ms$ demonstrating the ability to simulate a full-length TCV shot at 10kHz within minutes. We note that this performance is typically observed in practice and not specific to TCV \#61400.

\subsection{FGE vs. FGEL}
\label{sec:fgel_comp_subsec}
To compare the full non-linear evolution with the evolution dictated by the linear state-space~system~\eqref{eq:state_space_sys}, we analyze the diverted configuration in Anamak \#2 (see Appendix~\ref{sec:anamak}) equipped with a feedback controller on the radial and vertical plasma position as well as the plasma current. We perturb the equilibrium with a series of vertical kicks actuated by briefly applying a high voltage to one of the coils. The simulation is carried out over 100$ms$ in total with a temporal resolution of 100$\mu s$. The state-space system is computed from a linearization based on the initial equilibrium and evolved using implicit Euler stepping.

We observe a near exact match in the predicted dynamics between the full linear model and the linearized state-space representation for small deviations from the initial equilibrium. We observe that a significant mismatch arises only as the vertical displacement is sufficient to change the configuration from diverted to limited. In this case, the linearization of the initial diverted equilibrium does not accurately describe the dynamics of the limited equilibrium. This highlights the limits of the state-space model for cases with a significantly changing plasma configuration. We observe that the mismatch in the dynamics diminishes when the equilibrium is brought back close to its initial configuration. We report a computational time per time step of $10.5ms$ for the full non-linear solver and $0.9ms$ for the linearized system on an Intel i9-12900K CPU with the standard Anamak computational grid ($34{\times}33$).

\subsection{Growth Rate Comparison With KINX}
We compare growth rates of axisymmetric vertical resistive wall modes with the KINX code \citep{kinx_code_paper} which solves for eigenvalues in the variation of the total energy with plasma displacement in the ideal MHD setting. As KINX assumes a perfectly conducting plasma, we employ the 1D CDE with a very large conductivity to provide a fair comparison. We choose a set of eleven spline basis functions to represent the plasma profiles. These are specifically chosen to contain the degrees of freedom needed to represent non-zero values at the plasma boundary. This ensures that induced edge currents can be approximately modeled in the FGE growth rate computation. Extra steps were taken to ensure that both codes are computing the unstable modes with the same resistive wall representation. It should be noted that the comparison with KINX does not serve as a definite reference but as a useful benchmark with an established code with more fundamental modeling assumptions.
We compare four cases, a negative triangularity shot, a TCV ``standard'' shot, a shot with an X-point target and a fictitious doublet plasma which features two unstable vertical modes.

The results of this comparison are shown in Figure~\ref{fig:kinx_comp}. We observe a qualitative agreement with FGE and KINX detecting unstable modes with the same overall structure and the computed growth rates of the unstable modes largely agreeing. However, KINX tends to underestimate the instability compared to FGE for the negative triangularity shot and the doublet. Some disparities are expected as both codes solve very different systems. The larger differences might be due to KINX solving for the plasma response on a finer flux-coordinate grid and the different treatment of the vessel. We note that resolving induced currents close to the X-point does not only play a large role for the doublet plasma but also for single-axis plasmas and, together with the limited resolution of the 1D profiles, might add to the observed differences. However, we emphasize that FGE is also able to model the vertical modes with a more realistic conductivity profile where the plasma response in the cold edge plays a less significant role. we refer to \citep[Appendix D]{marchioni_thesis} for a more exhaustive comparison between vertical growth rate estimates from FGE and KINX.

\subsection{CDE Comparison with RAPTOR}
A comparison with the RAPTOR code is performed by computing a benchmark equilibrium evolution on Anamak \#1 (see Appendix~\ref{sec:anamak}) through coupling RAPTOR \citep{RAPTOR_code} with the FBT \citep{FBT_code} inverse Grad-Shafranov solver. To this end, an iteration is carried out between FBT computing a sequence of equilibria with prescribed shape, profiles and loop voltage and RAPTOR computing the profile evolution based on the geometric properties of these equilibria. After converging, the solution represents a fully consistent equilibrium and profile evolution. Finally, this is compared with the evolution as computed by FGE with the 1D CDE using the coil voltage traces computed by FBT. As the 1D CDE does not predict the evolution of kinetic profiles like temperature and density, we prescribe the pressure and conductivity profiles from RAPTOR to give a fair comparison.

Results computed for the fictitious circular plasma of Anamak \#1 with off-axis ECCD are shown in Figure~\ref{fig:raptor_comp_dynamic}. The simulation is run for $100ms$ with $100\mu s$ resolution. The traces show a good agreement between RAPTOR and \texttt{CDE-1D} for the profile evolution throughout the changing current distribution.

Additionally, we perform a comparison between stationary-state solutions for a diverted plasma with off-axis ECCD. The FBT-RAPTOR coupling is carried out with the RAPTOR native stationary-state solver \citep{Van_Mulders_raptorsteadystate} and compared to the FGE initialization computed with \texttt{static-CDE-1D}. Results are shown in Figure~\ref{fig:raptor_comp_static}. We see that the FGE solver is able to correctly reproduce the highly variable $TT'$ profile.

\subsection{Discharge Resimulation}
We showcase the capabilities of FGE resimulating TCV \#61400 from $t{=}0.1s$ to $t{=}1.8s$ with a temporal resolution of $100\mu s$ using the analytical jacobian computations and residual $F_{p1}$. The tolerance was chosen to be tolF=$10^{-8}$. For this test, we additionally constrain the plasma beta, as well as the internal inductance to match the reconstruction and run with the \texttt{OhmTor} 0D CDE to model the plasma current response. The vessel is modeled using $80$ eigenmodes with experimentally determined resistances. During the simulation, the plasma is controlled by the same controller as used in the experiment. 
It should be noted, however, that the digital controller implemented in the simulation does not perfectly replicate the analog controller from the experiment leading to a slightly worse performance for vertical stabilization and notable oscillations in the early phases. The results from the simulation are presented in Figure~\ref{fig:resim}, with experimental reference traces taken from the LIUQE reconstruction.

Albeit the significant changes in the plasma configuration throughout the shot, we observe an excellent agreement between the FGE simulation and the reconstructions from the experiment. Shaping as well as position of the plasma are nearly identical to the experiment. Using the same controller, the coil currents also behave very similar to their experimental counterparts which highlights the utility of FGE for controller development and validation. The full simulation ran in $225.9s$ on an Intel i9-12900K CPU which translates to ca. $13ms$ per time step. The slight increase in computational time compared to the results presented in Section~\ref{sec:convergence_studies} is likely due to noisy $\beta_p$ and $q_A$ references used to constrain the plasma profiles.

\section{Conclusion}
\label{sec:conclusion}
We have presented the Free-Boundary Grad-Shafranov Evolutive (FGE) code, a fast and control-oriented tool for simulating the time evolution of free-boundary tokamak equilibria. FGE self-consistently couples the Grad–Shafranov solution with external circuit dynamics and a range of current diffusion models, enabling rapid simulations suitable for control design and integrated scenario development. Numerical benchmarks verify the solver performance, demonstrate accurate reproduction of linearized plasma dynamics, and show good agreement with established codes such as RAPTOR and KINX. Furthermore, a resimulation of a highly dynamic TCV discharge confirms the code’s predictive capabilities under experimentally relevant conditions. By combining control‑oriented design with fast, reliable solvers and accurate physical models, FGE offers a comprehensive and effective framework for free-boundary plasma equilibrium simulation.

\FloatBarrier

\vspace{1em}
\noindent\sffamily\textbf{Acknowledgements}\normalfont\\
\small
The authors thank Dr. Jean-Marc Moret for the support in the early stages of the development of FGE and Cassandre Contré for help with the RAPTOR-FBT coupling. This work has been carried out within the framework of the EUROfusion Consortium, via the Euratom Research and Training Programme (Grant Agreement No 101052200 — EUROfusion) and funded by the Swiss State Secretariat for Education, Research and Innovation (SERI). Views and opinions expressed are however those of the author(s) only and do not necessarily reflect those of the European Union, the European Commission, or SERI. Neither the European Union nor the European Commission nor SERI can be held responsible for them.
\normalsize

\bibliography{Bibliography}

\section{Appendix}

\subsection{Plasma Constraint Details}
\label{sec:appendic_agcon}
In this section, we give a comprehensive list of the available choices for the $a_g$ residual function $\Lambda$ introduced in equation~\eqref{eq:constraints}. We note that although $\Lambda$ takes as input the $a_g$ coefficients together with other plasma parameters, we omit these other dependencies for readability and to highlight the role of constraining the $a_g$ degrees of freedom.

To clearly express the implemented residuals, we formally introduce $\mathcal{Y}$ as the index set underlying the $(R,Z)$-grid, on which the discretized toroidal current distribution $I_y$ is defined, and $\Psi_y$ as the discretized representation of $\psi$ on the same grid. We further define $\mathcal{Y}_\mathcal{P} \subseteq \mathcal{Y}$ as subset of indices lying inside the plasma domain.

\begin{itemize}
    \item \textbf{Direct constraints of the $\boldsymbol{a_g}$ coefficients.} Any one of the $a_g$ coefficients can be directly constrained to some reference value:
        \begin{equation}
            \Lambda_{a_g}(a_g) = a_g - a_g^{\text{ref}}, \qquad g \in \mathcal{G}_p \cup \mathcal{G}_T.
        \end{equation}
    \item \textbf{Plasma current $\boldsymbol{I_p}$.} The plasma current is constrained by the residual
    \begin{equation}
    \label{eq:ip_constraint}
    \Lambda_{I_p}(a_g) = I_p^{\text{ref}} - \sum_{i \in \mathcal{Y}} (I_y(a_g))_i,
    \end{equation}
    where the dependence of $I_y$ on the $a_g$ coefficients is given by the $T_{yg}$ matrix as described in Section~\ref{sec:equilibrium_equations}.

    \item \textbf{Poloidal beta $\boldsymbol{\beta_p}$.}
    We start from the expression for $\beta_p$ from \citep{Lao_1985_betap}
    \begin{equation}
        \beta_p = \frac{\int_{V_\mathcal{P}} p \dx{V}}{V \mu_0 I_p^2 / 2 l^2}.
    \end{equation}
    With the approximations $V\approx 2 \pi^2 \kappa a^2 R_0$, $l \approx 2 \pi \sqrt{\kappa} a$ for the plasma volume and the poloidal perimeter of the plasma, respectively, we arrive at the expression
    \begin{align}
        \beta_p = \frac{\sum_{i \in \mathcal{Y}} \sum_{g \in \mathcal{G}_p} 2 \pi R_i \, a_g\ g((\Psi_y)_i)}{\frac{1}{4} \mu_0 R_0 I_p^2}.
    \end{align}

    Defining the plasma current by
    \begin{equation}
    I_p(a_g) \coloneqq \sum_{i \in \mathcal{Y}} (I_y(a_g))_i,
    \end{equation}
    dependent on the $a_g$ coefficients, the constraint reads
    \begin{equation}
    \begin{aligned}
        \Lambda_{\beta_p}(a_g) = \sum_{i \in \mathcal{Y}} \sum_{g \in \mathcal{G}_p} 2 \pi R_i \, a_g\ g((\Psi_y)_i)\\
        - \beta_p^{\text{ref}} \left(\frac{1}{4} \mu_0 R_0 I_p^2(a_g)\right).
    \end{aligned}
    \end{equation}

    \item \textbf{Toroidal beta $\boldsymbol{\beta_t}$.}
    By expressing the toroidal magnetic pressure as toroidal magnetic energy divided by volume $W_{tor} / V$, the definition of toroidal beta reads
    \begin{equation}
        \beta_t = \frac{\int_{V_\mathcal{P}} p \dx{V}}{W_{\text{tor}}}.
    \end{equation}
    Approximating $W_{\text{tor}}$ by its vacuum component
    \begin{equation}
        W_{\text{tor}} \approx \int_{V_\mathcal{P}} \frac{B^2_{\phi, \text{vac}}}{2 \mu_0} \dx{V} \approx \frac{\pi R_0^2 B^2_{\phi,0}}{\mu_0} 
        \sum_{i \in \mathcal{Y}_\mathcal{P}} \frac{1}{R_i}
    \end{equation}
    yields the constraint residual
    \begin{equation}
        \Lambda_{\beta_t}(a_g) = \sum_{i \in \mathcal{Y}_\mathcal{P}} \sum_{g \in \mathcal{G}_p} 2 \pi R_i \, a_g g((\Psi_y)_i) - \beta_t^{\text{ref}} W_{\text{tor}}.
    \end{equation}

    \item \textbf{Normalized internal inductance $\boldsymbol{\ell_i}$.}
    In MEQ, the $\ell_i(3)$ \citep{jackson_li} expression for the normalized internal inductance is used:
    \begin{equation}
        \ell_i = \frac{4 W_p}{\mu_0 R_0 I_p^2}.
    \end{equation}
    The poloidal magnetic energy is computed using the finite difference representation of $\abs{\nabla \Psi}^2$:
    \begin{equation}
        W_p = \int_{V_\mathcal{P}} \frac{B_p^2}{2 \mu_0} \dx{V} = \frac{1}{4 \pi \mu_0}\sum_{i \in \mathcal{Y}_\mathcal{P}} \frac{\abs{\nabla \Psi}_i^2}{R_i}.
    \end{equation}
    Including the dependence of $I_p$ on the $a_g$ coefficients, the constraint reads
    \begin{equation}
        \Lambda_{\ell_i}(a_g) = W_p - \ell_i^{\text{ref}} \left(\frac{1}{4} \mu_0 R_0 I_p^2(a_g)\right).
    \end{equation}

    \item \textbf{Safety factor at magnetic axis $\boldsymbol{q_A}$.}
    Following \citep{liuqe_paper}, the safety factor at the axis is expressed as
    \begin{equation}
        q_A = \frac{\abs{\tr H }}{\sqrt{\det H}} \frac{T_A}{\mu_0 R_A^2 j_{\phi,A}},
    \end{equation}
    where $H$ is the Hessian of $\psi$ at the axis and $T_A$ is the value of the $T$-profile at the magnetic axis approximated using the vacuum component of the toroidal magnetic field $B_{\phi, 0}$ at $R_0$:
    \begin{equation}
        T_A \approx R_0 B_{\phi,0} + \frac{1}{R_0 B_{\phi,0}} \sum_{g \in \mathcal{G}_T} a_g\ g(\psi_A).
    \end{equation}
    With the toroidal current density at the magnetic axis being given by
    \begin{equation}
        j_{\phi,A} = R_A \sum_{g \in \mathcal{G}_p} a_g \ g'(\psi_A) + \frac{1}{R_A} \sum_{g \in \mathcal{G}_T} a_g \ g'(\psi_A),
    \end{equation}
    we arrive at the constraint residual
    \begin{equation}
        \Lambda_{q_A}(a_g) = q_A^{\text{ref}} \left( \mu_0 R_A^2 \frac{\sqrt{\det H}}{\abs{\tr H}}\right) j_{\phi,A} - T_A.
    \end{equation}
    
\end{itemize}

\subsection{CDE derivations and relations}
\label{sec:cde_derivs}
In this section we outline the derivations for the different CDE formulations as well as viewing them in the context of conservation laws.

Many of the CDE formulations in Section~\ref{sec:cdes} solely present a model for the evolution of toroidal currents in the plasma. Consequently, their derivation begins with the toroidal projection of Ohm's Law:
\begin{equation}
\label{eq:toroidal_ohms_law}
    \sigma E_\phi = j_\phi - j_{ni, \phi}.
\end{equation}
While only the $\bfB$-parallel formulation of Ohm's law is physically accurate, adopting the toroidal projection greatly simplifies the CDE equations by omitting terms related to poloidal currents. In most practical cases, this approximation remains close to the $\bfB$-parallel form, with deviations often being negligible, particularly when parameters such as the bulk plasma resistance $R_p$ are empirically determined.

Furthermore, non-inductive currents are most often assumed to be driven in the $\bfB$-parallel direction and many transport codes represent them in terms of the flux surface averaged quantity $\langle \bfj_{ni} \cdot \bfB \rangle / B_0$ \citep{RAPTOR_code, astra}. In contrast, all CDEs discussed in Section~\ref{sec:cdes}, with the exception of \texttt{CDE‑1D} and \texttt{CDE-stationary-state}, consider only their toroidal contribution. As we will see later, the total non-inductive current $I_{ni}$ used in many 0D formulations, when considered in the context of the bulk plasma resistivity $R_p$ defined below, does not directly correspond to the integral of the toroidal component of $\bfj_{ni}$ and should be treated with care.

Multiplying the toroidal projection of Ohm's Law with $j_\phi$ and dividing by the conductivity yields a local power balance equation:
\begin{equation}
\label{eq:toroidal_power_balance}
    j_\phi E_\phi = \frac{1}{\sigma} j_\phi (j_\phi - j_{ni,\phi}).
\end{equation}
Here, the left hand side represents work done on toroidal currents in the plasma and the right hand side describes the Ohmic dissipation. Many CDEs can be interpreted as integral formulations of this equation under different assumptions and approximations. Furthermore, this expression lends itself to be reformulated in an energy conservation form as the dissipated energy must come from the plasma or external conductors and actuators.

The \texttt{OhmTor} CDE is derived from equation~\eqref{eq:toroidal_power_balance} by integrating and expressing the toroidal electric field using the time derivative of the poloidal flux
\begin{align}
    -\int_{\mathbb{H}} j_\phi \frac{\partial \psi}{\partial t} \dx{R} \dx{Z} = \int_{\mathbb{H}} \frac{2 \pi R}{\sigma} j_\phi (j_\phi - j_{ni,\phi})  \dx{R} \dx{Z}.
\end{align}
Here, the integral over the whole space is done by multiplying by $2 \pi R$ and integrating over the right half plane $\mathbb{H}$.
Defining the bulk plasma resistivity as
\begin{equation}
\label{eq:bulk_resistivity}
    R_p = \frac{1}{I_p^2} \int_{\mathbb{H}} \frac{2 \pi R}{\sigma} j_\phi^2  \dx{R} \dx{Z}
\end{equation}
yields
\begin{equation}
    0 = \frac{1}{I_p} \int_{\mathbb{H}} j_\phi \frac{\partial \psi}{\partial t} \dx{R} \dx{Z} + R_p (I_p - I_{ni}),
\end{equation}
where $I_{ni}$ has to be defined accordingly.
Discretizing the integral as a sum over grid cells finally results in the \texttt{OhmTor} formulation.

This expression can be further modified to include changes in the plasma magnetic energy by decomposing $\psi$ into the plasma component $\psi_{plasma}$ and the component $\psi_{ext}$ arising from currents in external conductors:
\begin{align*}
    0 &= \int _{\mathbb{H}}j_\phi \frac{\partial \psi}{\partial t}  \dx{R} \dx{Z} + R_p I_p (I_p - I_{ni})\\
    &= \int_{\mathbb{H}} j_\phi \frac{\partial \psi_{plasma}}{\partial t} \dx{R} \dx{Z} + \int_{\mathbb{H}} j_\phi \frac{\partial \psi_{ext}}{\partial t}  \dx{R} \dx{Z} \\
    &\qquad \qquad\qquad\qquad\qquad\qquad\qquad+ R_p I_p (I_p - I_{ni}).
\end{align*}
Using equation~\eqref{eq:GS} and partial integration:
\begin{align*}
    &\int_\mathbb{H} j_\phi \frac{\partial \psi_{plasma}}{\partial t} \dx{R} \dx{Z}\\
    &= -\int_{\mathbb{R}^3}  \frac{1}{4 \pi^2 R^2 \mu_0} \Delta^{\ast}\psi_{plasma}\frac{\partial \psi_{plasma}}{\partial t} \dx{V}\\
    &= \int_{\mathbb{R}^3}  \frac{1}{4 \pi^2 R^2 \mu_0} \nabla \psi_{plasma} \cdot \frac{\partial(\nabla \psi_{plasma})}{\partial t} \dx{V}\\
    &= \int_{\mathbb{R}^3}  \frac{\bfB_{p,plasma}\! \cdot\! \dot{\bfB}_{p,plasma}}{\mu_0} \dx{V} \\
    &= \frac{\partial}{\partial t} \left(\int_{\mathbb{R}^3}  \frac{B_{p,plasma}^2}{2 \mu_0} \dx{V} \right)\! ,
\end{align*}
results in
\begin{equation}
\label{eq:ohmtor_power_balance}
\begin{aligned}
    0 &= \int _{\mathbb{H}}j_\phi \frac{\partial \psi}{\partial t}  \dx{R} \dx{Z} + R_p I_p (I_p - I_{ni})\\
    &= \frac{\partial}{\partial t} \left(\int_{\mathbb{R}^3} \frac{B_{p,plasma}^2}{2 \mu_0} \dx{V} \right) + \int_{\mathbb{H}} j_\phi \frac{\partial \psi_{ext}}{\partial t}  \dx{R} \dx{Z}\\
    &\qquad\qquad\qquad\qquad\qquad \qquad\qquad + R_p I_p (I_p - I_{ni}).
\end{aligned}
\end{equation}
Hence, \texttt{OhmTor} describes a power balance equation relating the dissipated energy to the change in the magnetic energy of the plasma as well as the work done by external coils. By defining the self-inductance of the plasma as
\begin{equation}
L_p = \frac{1}{I_p^2} \int_{\mathbb{R}^3}  \frac{B_{p,plasma}^2}{\mu_0} \dx{V},
\end{equation}
we arrive at
\begin{equation}
0= \frac{\partial}{\partial t} \left(\frac{1}{2} L_p I_p^2\right) + \int_{\mathbb{H}}  j_\phi \frac{\partial \psi_{ext}}{\partial t}  \dx{R} \dx{Z} + R_p I_p (I_p - I_{ni}).
\end{equation}
Assuming a the plasma to be a rigid conductor with constant self-inductance and dividing by $I_p$ yields the \texttt{OhmTor-rigid} formulation. As mentioned in Section~\ref{sec:cdes}, \texttt{OhmTor-rigid} offers the option to recompute $L_p$ dynamically. In this case, the residuals of \texttt{OhmTor-rigid} and \texttt{OhmTor} differ only by the term $\frac{1}{2}\dot{L}_p I_p$:
\begin{equation}
\begin{aligned}
    &\frac{1}{I_p} I_y^T \frac{\partial \psi}{\partial t} + R_p (I_p - I_{ni})\\
    &= L_p \dot{I}_p + M_{pe} \dot{I}_e + R_p (I_p - I_{ni}) + \frac{1}{2}\dot{L}_p I_p.
\end{aligned}
\end{equation}

Closely related is a CDE commonly found under the name \textit{Ejima equation} \citep{Ejima_1982}, which takes the form
$$0 = \frac{1}{I_p} \frac{\partial}{\partial t} \left(\frac12 L_i I_p^2 \right) + \frac{\partial \psi_{B}}{\partial t} + R_p (I_p - I_{ni}),$$
where $\psi_B$ denotes the flux at the plasma boundary.
While this equation is currently not implemented in FGE, we nonetheless include it in this dicussion for reference.
To arrive at this equation involving the internal inductance, one integrates equation~\eqref{eq:toroidal_power_balance} over the plasma volume $V_\mathcal{P}$ instead of the whole space.
\begin{equation}
\label{eq:internal_inductance_power_balance}
\begin{aligned}
    0 &= \int_{V_\mathcal{P}} j_\phi \frac{1}{2 \pi R} \frac{\partial \psi}{\partial t} \dx{R} \dx{Z} + R_p I_p (I_p - I_{ni})\\
    &= \frac{\partial}{\partial t} \left(\int_{V_\mathcal{P}} \frac{B_{p}^2}{2 \mu_0} \dx{V} \right) - \int_{\partial V_\mathcal{P}} \frac{B_p^2}{2 \mu_0} v_b \cdot \dx{S}\\
    &- \int_{\partial V_\mathcal{P}} \frac{1}{4\pi^2 R^2 \mu_0} \frac{\partial \psi}{\partial t} \nabla \psi\! \cdot\! \dx{S} + R_p I_p (I_p - I_{ni}),
\end{aligned}
\end{equation}
where $v_b$ is the velocity of the plasma boundary.
In this case, integrating by parts introduces a boundary term but does not require the decomposition of the poloidal flux into the plasma and external contribution as all external currents are outside of the plasma domain. Furthermore, an additional term arises from switching the time derivative and the integral when integrating over the possibly moving plasma domain.

This way of expressing the power balance lends itself to an interpretation in the context of Poynting's theorem as the boundary term in the equation above is equal to the boundary integral of the Poynting vector:
\begin{equation}
\int_{\partial V_\mathcal{P}} \frac{1}{4\pi^2 R^2 \mu_0} \frac{\partial \psi}{\partial t} \nabla \psi\! \cdot\! \dx{S} = -\int_{\partial V_\mathcal{P}} \frac{1}{\mu_0}(\bfE_\phi \times \bfB_p) \cdot \dx{S}.
\end{equation}
In short, the power available to be dissipated comes from the change of internal energy with an added term due to the moving boundary and the electromagnetic energy entering the plasma volume.
Assuming no shape changes, we get $v_b=0$ and $\partial_t\psi$ constant on the boundary
\begin{align}
    -\int_{\partial V_\mathcal{P}} \frac{1}{4\pi^2 R^2 \mu_0} \frac{\partial \psi}{\partial t} \nabla \psi\! \cdot\! \dx{S}  &= \frac{\partial \psi_{B}}{\partial t} \frac{1}{\mu_0}\int_{\partial A_\mathcal{P}} \bfB_p \cdot \dx{\ell}\notag \\
    &= \frac{\partial \psi_{B}}{\partial t} I_p
\end{align}
and the Ejima equation is recovered. A different path for the derivation involves starting from the \texttt{OhmTor} equation and decomposing the change of the total magnetic energy into the change of internal energy and the energy flux across the boundary.

If shape changes are present, the Ejima equation differs from equation~\eqref{eq:ohmtor_power_balance} only in a term involving boundary movement:
\begin{equation}
\begin{aligned}
    &\frac{1}{I_p}\int _{\mathbb{H}}j_\phi \frac{\partial \psi}{\partial t}  \dx{R} \dx{Z} + R_p (I_p - I_{ni})\\
    &= \frac{1}{I_p} \frac{\partial}{\partial t} \left(\frac12 L_i I_p^2 \right) + \frac{\partial \psi_{B}}{\partial t} + R_p (I_p - I_{ni})\\
    &+ \frac{1}{2 \mu_0 I_p} \int_{\partial A_{\mathcal{P}}}\left(\frac{\partial \psi}{\partial t} - \frac{\partial \psi_B}{\partial t}\right) B_p \cdot \dx{\ell}
\end{aligned}
\end{equation}
This follows from the identity  $v_b = (\frac{\partial \psi_B}{\partial t} - \frac{\partial \psi}{\partial t}) \frac{\nabla \psi}{|\nabla \psi|^2}$:
\begin{align*}
    &\int_{\partial V_\mathcal{P}} \frac{1}{4\pi^2 R^2 \mu_0} \frac{\partial \psi}{\partial t} \nabla \psi\! \cdot\! \dx{S} + \int_{\partial V_\mathcal{P}} \frac{B_p^2}{2 \mu_0} v_b \cdot \dx{S}\\
    &=\int_{\partial V_\mathcal{P}} \frac{1}{4\pi^2 R^2 \mu_0} \frac{\partial \psi}{\partial t} \nabla \psi\! \cdot\! \dx{S} + \int_{\partial V_\mathcal{P}} \frac{|\nabla \psi|^2}{8 \pi^2 R^2 \mu_0} v_b \cdot \dx{S}\\
    &= \int_{\partial V_\mathcal{P}}\frac{1}{8\pi^2 R^2 \mu_0} \left(\frac{\partial \psi}{\partial t} + \frac{\partial \psi_B}{\partial t} \right) \nabla \psi \cdot \dx{S}\\
    &= -\int_{\partial A_\mathcal{P}} \frac{1}{2 \mu_0} \left(\frac{\partial \psi}{\partial t} + \frac{\partial \psi_B}{\partial t} \right) B_p \cdot \dx{\ell}\\
    &= -\int_{\partial A_\mathcal{P}} \frac{1}{2 \mu_0} \left(\frac{\partial \psi}{\partial t} - \frac{\partial \psi_B}{\partial t} \right) B_p \cdot \dx{\ell}- I_p \frac{\partial \psi_B}{\partial t}.
\end{align*}
This shows that while the term involving the plasma surface velocity $v_b$ in equation~\eqref{eq:internal_inductance_power_balance} can be recast to closely resemble the Poynting vector surface integral, these terms unfortunately do not cancel exactly to yield $\partial_t \psi_B\, I_p$.

As mentioned in Section~\ref{sec:cdes}, versions of the current diffusion equation based on flux surface averages are not easily applicable to the cartesian grid of FGE. However, \texttt{CDE-stationary-state} presents an exception and its derivation shares many steps with CDEs implemented in RAPTOR \citep{RAPTOR_code} or ASTRA \citep{astra}. The equation
\begin{equation}
\left. \frac{\partial \sPhi}{\partial t} \right|_\psi = \frac{\partial V}{\partial \psi} \langle \bfE \cdot \bfB\rangle
\end{equation}
from \citep{toroidal_flux_reference} presents the starting point. The parallel formulation of Ohm's law is used to give
\begin{equation}
\label{eq:CDE_flux_diff_starting_point}
\begin{aligned}
    \sigma_\| \left. \frac{\partial \sPhi}{\partial t} \right|_\psi = &\frac{\partial V}{\partial \psi} \langle \bfj \cdot \bfB \rangle - \frac{\partial V}{\partial \psi} \langle \bfj_{ni} \cdot \bfB \rangle = \\
    &\frac{\partial V}{\partial \psi} \langle j_\phi B_\phi \rangle + \frac{\partial V}{\partial \psi} \langle \bfj_p \cdot \bfB_p\rangle  - \frac{\partial V}{\partial \psi} \langle \bfj_{ni} \cdot \bfB \rangle.
\end{aligned}
\end{equation}
The right hand side is expressed using
\begin{align}
    \langle j_\phi B_\phi \rangle &= T \left\langle \frac{j_\phi}{R}\right\rangle = 2\pi T \frac{\partial \psi}{\partial V} \int_{\partial A_\psi} j_\phi \frac{\dx{\ell}}{\abs{\nabla \psi}}\notag \\
    &= 2\pi T \frac{\partial \psi}{\partial V} \frac{\partial I_p(\psi)}{\partial \psi},\\
    \langle \bfj_p \cdot \bfB_p\rangle &= \frac{1}{\mu_0} \langle (\nabla \! \times\! \bfB_t) \cdot \bfB_p \rangle = -\frac{2 \pi}{\mu_0} T' \langle \bfB_p \cdot \bfB_p\rangle\notag \\
    &= -\frac{2 \pi}{\mu_0} T' \frac{\partial \psi}{\partial V} \int_{\partial A_\psi} B_p \dx{\ell} = -\frac{2 \pi}{\mu_0} T' \frac{\partial \psi}{\partial V} I_p(\psi).
\end{align}
Here, $I_p(\psi)$ denotes the toroidal current contained inside the flux surface with label $\psi$. Together, this yields
\begin{align}
    \sigma_\| \left. \frac{\partial \sPhi}{\partial t} \right|_\psi &= 2\pi T \frac{\partial I_p(\psi)}{\partial \psi} - \frac{2 \pi}{\mu_0} T' I_p(\psi) - \frac{\partial V}{\partial \psi} \langle \bfj_{ni} \cdot \bfB \rangle\notag \\
    &= 2 \pi T^2 \frac{\partial}{\partial \psi} \left(\frac{I_p(\psi)}{T} \right) - \frac{\partial V}{\partial \psi} \langle \bfj_{ni} \cdot \bfB \rangle.
\end{align}
Dividing by $T^2$ and integrating over the poloidal flux coordinate from the axis $\psi_A$ to the plasma boundary $\psi_B$ gives
\begin{align}
    0 = &-\int_{\psi_A}^{\psi_B} \frac{\sigma_\|}{T^2} \left. \frac{\partial \sPhi}{\partial t}\right|_\psi \dx{\psi} + 2 \pi \int_{\psi_A}^{\psi_B} \frac{\partial}{\partial \psi} \left(\frac{I_p(\psi)}{T}\right) \dx{\psi}\notag \\
    &-\int_{\psi_A}^{\psi_B} \frac{1}{T^2} \frac{\partial V}{\partial \psi} \langle \bfj_{ni} \cdot \bfB \rangle \dx{\psi}\notag \\
    = &-\int_{0}^{\sPhi_B} \frac{\sigma_\|}{T^2} \left(\frac{\partial \sPhi(\psi)}{\partial t} -  \frac{\partial \psi}{\partial t}\right)\dx{\sPhi} + 2\pi \frac{I_p - I_{ni}}{T_B},
\end{align}
where $I_{ni}$ has to be defined accordingly.
Under the stationary-state assumption, $\partial_t\sPhi(\psi) = 0$ and $\partial_t \psi = const. = \frac12 (\partial_t \psi_A + \partial_t \psi_B)$ resulting in the \texttt{CDE-stationary-state} residual.

Regarding the derivation of \texttt{CDE-1D}, as well as the discussion of its associated power balance equations, we directly refer to \citep{heumann_weak_formulation}. Its simplified version, \texttt{CDE-OhmTor-1D}, can be viewed as an approximation of \texttt{CDE-1D} without poloidal currents ($T'=0$) or as the product of directly integrating the toroidal projection of Ohm's Law against test functions. As important terms relating to the power balance of poloidal currents from \texttt{CDE-1D} are left out, \texttt{CDE-OhmTor-1D} doesn't lend itself to a simple representation in terms of energy conservation. We note that \texttt{CDE-OhmTor-1D} could be turned into a power balance equation by multiplying by $j_\phi$ before integrating, albeit with the downside of increasing the complexity of the residual.

In the above derivations, the definition of the total non-inductive current is not necessarily intuitive. For the \texttt{OhmTor}-type CDEs with the bulk plasma resistivity defined as in equation~\eqref{eq:bulk_resistivity}, $I_{ni}$ is formally given by
\begin{align}
    I_{ni} = I_p \left(\frac{\int_\mathbb{H} \frac{2 \pi R}{\sigma} j_\phi j_{ni,\phi} \dx{R} \dx{Z}}{\int_\mathbb{H} \frac{2 \pi R}{\sigma} j_\phi^2 \dx{R} \dx{Z}}\right)\! ,
\end{align}
which can be interpreted as a fraction of $I_p$ based on a weighted average of the non-inductive current density.

For \texttt{CDE-stationary-state}, the appropriate definition of the total non-inductive current reads
\begin{align}
    I_{ni} = \int_{\psi_A}^{\psi_B} \frac{T_B^2}{T^2} \left(\frac{1}{2 \pi R_0}  \frac{\partial V}{\partial \psi}\right) \frac{\langle \bfj_{ni} \cdot B \rangle}{B_0} \dx{\psi},
\end{align}
which, assuming a small diamagnetic contribution, translates to an approximation of the integral of the non-inductive current density over the poloidal cross-section of the plasma.

While these definitions are unwieldy, they correctly correspond to the total non-inductive current in an important edge case. If the whole plasma current is supported by non-inductive current drive and steady-state is reached, equation~\eqref{eq:toroidal_ohms_law}
implies $j_\phi = j_{ni,\phi}$ and equation~\eqref{eq:CDE_flux_diff_starting_point} implies $\langle \bfj \cdot \bfB \rangle = \langle \bfj_{ni} \cdot \bfB \rangle$. Therefore, for both the \texttt{OhmTor}-type CDEs and \texttt{CDE-stationary-state}, the respective definitions of the total non-inductive current correctly satisfy $I_{ni} = I_p$ in the case of a fully non-inductively driven steady-state plasma.
However, in general, the notion of $I_{ni}$ in 0D CDEs carries some caveats as the differences arising from different deposition locations in the plasma cannot be resolved. Hence, 1D CDEs are paramount when a consistent treatment of non-inductive currents is required.

\subsection{Anamak}
\label{sec:anamak}
MEQ implements a ficticious tokamak ``Anamak'' used as a toy example for code development, validation and benchmarking. Anamak features an elliptical vacuum vessel with active coils consisting of single windings being distributed uniformly around. 
Equations for the limiter, vessel and coil positions are given by
\begin{align*}
    \begin{pmatrix}
        R_l\\
        Z_l
    \end{pmatrix}
    &= \begin{pmatrix}
        R_0\\
        0
    \end{pmatrix} +
    a_l \begin{pmatrix}
        \cos(\theta)\\
        \kappa \sin(\theta)
    \end{pmatrix},\\
    \begin{pmatrix}
        R_v\\
        Z_v
    \end{pmatrix}
    &= \begin{pmatrix}
        R_0\\
        0
    \end{pmatrix} +
    a_v \begin{pmatrix}
        \cos(\theta)\\
        \kappa \sin(\theta)
    \end{pmatrix},\\
    \begin{pmatrix}
        R_c\\
        Z_c
    \end{pmatrix}
    &= \begin{pmatrix}
        R_0\\
        0
    \end{pmatrix} +
    a_c \begin{pmatrix}
        \cos(\theta_c)\\
        \kappa \sin(\theta_c)
    \end{pmatrix}.
\end{align*}
The limiter, vessel, and coil surfaces are defined using minor radii $a_l$, $a_v$, and $a_c$, typically $a_l = 0.45$, $a_v = 0.5$, and $a_c = 1.4 a_v$ are used. $\kappa$ denotes the elongation of the vessel and can be adjusted by the user. The angular coordinate $\theta$ is discretized into evenly spaced points from $0$ to $2\pi$. Similarly, $\theta_c$ spans $0$ to $2\pi$ with a number of points equal to the specified number of PF coils.

\begin{figure}[!h]
    \centering
    \includegraphics[width=\linewidth]{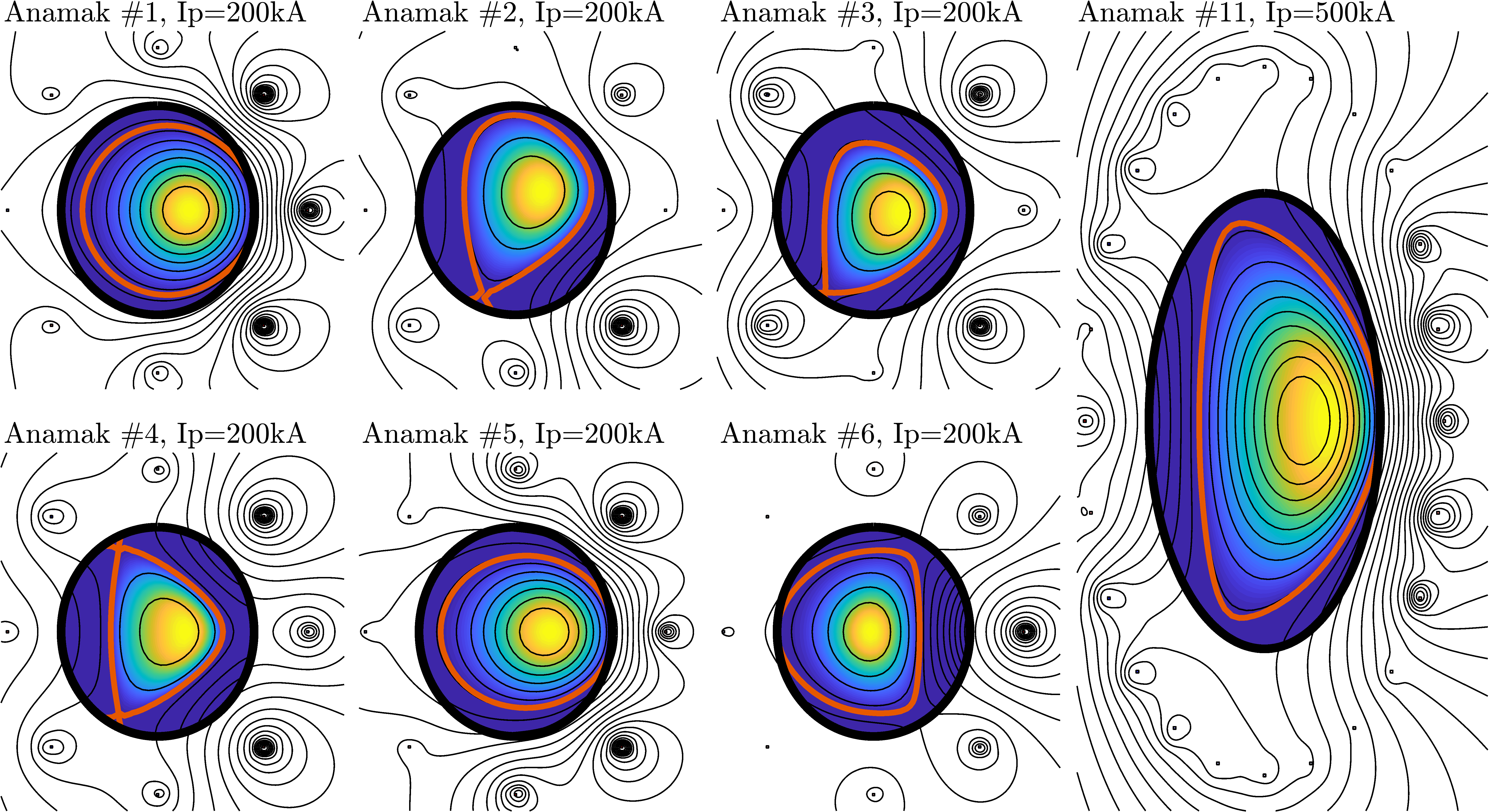}
    \caption{Plots of seven Anamak shots with different geometries. The poloidal flux is shown as contours where the LCFS is highlighted in red. The colormap shows the toroidal current distribution.}
    \label{fig:anamak_equis}
\end{figure}

This parametrization allows for easy changes in major radius, minor radius, vessel elongation, toroidal magnetic field and the number of active coils. A list of Anamak ``shots'' with standard properties is natively implemented in MEQ. These feature a circular plasma, a lower single null D-shaped plasma, a highly elongated plasma, a negative triangularity plasma, a doublet and more. Figure~\ref{fig:anamak_equis} shows a small selection of implemented Anamak equilibria.

\subsection{Analytical Examples for CDE Validation}
We detail analytical expressions for two circular equilibria in the low aspect ratio limit that can be used for validating 1D CDE formulations. The equilibria fulfill the cylindrical limit of equation~\ref{eq:GS}:
\begin{equation}
    \Delta \psi = -4 \pi^2 \left(\mu_0 R_0^2 p'(\psi) + TT'(\psi)\right)
\end{equation}
with $R_0 = 1$, as well as the B-parallel formulation of Ohm's law~\eqref{eq:ohmslaw}. We lay out two cases, one with inductive current drive and one with a decaying plasma current.

To this end, let $J_0 \colon \RR_+ \to \RR$, $J_1 \colon \RR_+ \to \RR$ be the Bessel functions of the first kind of orders zero and one, respectively. Furthermore, let $r_{0} = 2.404\ldots$ be the first root of $J_0$. Let $r \coloneqq |(R-R_0, Z-Z_0)|$ be the minor radial coordinate and $r_B>0$ be the minor radius of the plasma separating the plasma region ($r\leq r_B$) and the vacuum region ($r>r_B$). Let $B_{\phi,0}$ be the vacuum toroidal field strength at $R_0$ and $\alpha \coloneqq \frac{r_0}{r_B}$, $\beta, \gamma > 0$ be parameters such that $4 \pi^2 (\mu_0 \gamma + \beta^2) = \alpha^2$. Finally, let $\sigma_0>0$ be the bulk conductivity of the plasma.

The first equilibrium with inductive current drive is then defined by the following radial profiles:
\begin{equation}
    \begin{split}
        \psi(r) &\coloneqq
        \begin{cases}
            J_0(\alpha r) - \frac{t}{\mu_0 \sigma_0} & \quad \text{for}\ r \leq r_B,\\
            -\alpha J_1(r_B) r_B \ln\left(\frac{r}{r_B}\right) - \frac{t}{\mu_0 \sigma_0} & \quad \text{for}\ r > r_B,
        \end{cases}\\
        j_\phi(r) &\coloneqq \begin{cases}
        \frac{\alpha^2}{2 \pi \mu_0} J_0(\alpha r)  & \quad \text{for}\ r \leq r_B,\\
        0 & \quad \text{for}\ r > r_B,
        \end{cases}\\
        T(r) &\coloneqq \begin{cases}
            \beta (\psi(r) + \frac{t}{\mu_0 \sigma_0}) + B_{\phi, 0}  & \quad \text{for}\ r \leq r_B,\\
            B_{\phi, 0} & \quad \text{for}\ r > r_B,
        \end{cases}\\
        p'(r) &\coloneqq \begin{cases}
            \gamma (\psi(r) + \frac{t}{\mu_0 \sigma_0}) - \frac{\beta}{\mu_0}B_{\phi, 0}  & \quad \text{for}\ r \leq r_B,\\
            0 & \quad \text{for}\ r > r_B,
        \end{cases}\\
        \sigma_\|(r) &\coloneqq \begin{cases}
            \sigma_0 \alpha^2 \left(J_0(\alpha r) + \beta \frac{J_1^2(\alpha r)}{T(r)} \right)& \quad \text{for}\ r \leq r_B,\\
            0 & \quad \text{for}\ r > r_B.
        \end{cases}
    \end{split}
\end{equation}

For the same parameter definitions, given the time scale $\kappa \coloneqq \alpha^2 / \mu_0 \sigma_0$, the second equilibrium with exponentially decaying plasma current is given by:
\begin{equation}
    \begin{split}
        \psi(r) &\coloneqq \begin{cases}
        J_0(\alpha r) e^{-\kappa t} & \quad \text{for}\ r \leq r_B,\\
        -\alpha J_1(r_B) r_B \ln\left(\frac{r}{r_B}\right) e^{-\kappa t} & \quad \text{for}\ r > r_B,
        \end{cases}\\
        j_\phi(r) &\coloneqq \begin{cases}
        \frac{\alpha^2}{2 \pi \mu_0} \psi  & \quad \text{for}\ r \leq r_B,\\
        0 & \quad \text{for}\ r > r_B,
        \end{cases}\\
        T(r) &\coloneqq \begin{cases}
            \beta \psi(r) + B_{\phi, 0}  & \quad \text{for}\ r \leq r_B,\\
            B_{\phi, 0} & \quad \text{for}\ r > r_B,
        \end{cases}\\
        p'(r) &\coloneqq \begin{cases}
            \gamma \psi(r) - \frac{\beta}{\mu_0} B_{\phi, 0}  & \quad \text{for}\ r \leq r_B,\\
            0 & \quad \text{for}\ r > r_B.
        \end{cases}\\
        \sigma_\|(r) &\coloneqq \begin{cases}
            \sigma_0& \quad \text{for}\ r \leq r_B,\\
            0 & \quad \text{for}\ r > r_B.
        \end{cases}
    \end{split}
\end{equation}

\begin{figure}[!t]
    \centering
    \includegraphics[width=\linewidth]{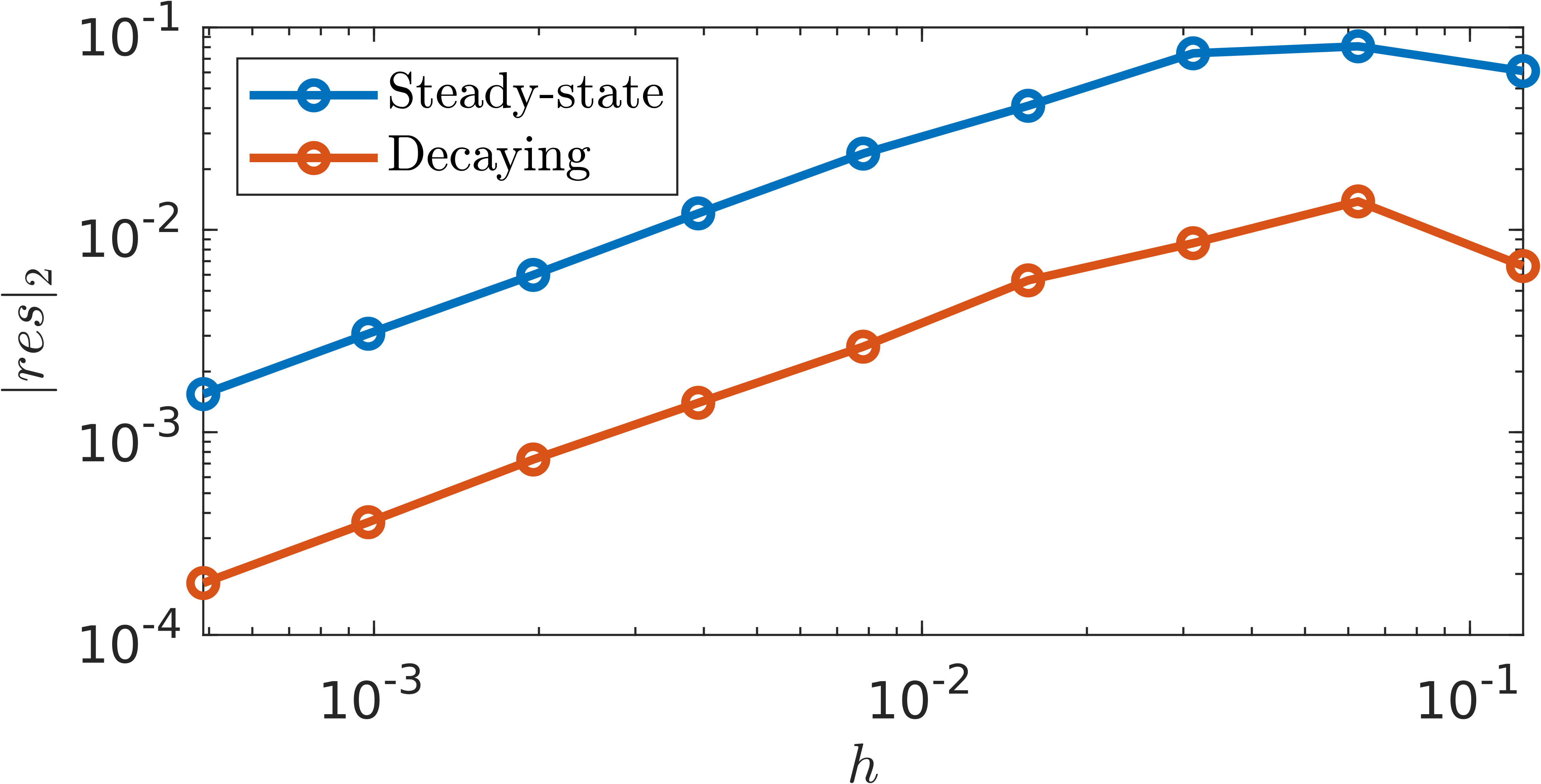}
    \caption{Benchmark of the normalized \texttt{CDE-1D} residual on the analytical screw-pinch configurations. $h$ is the fidelity parameter with the number of grid cells given by $N_x = \frac{1}{h^2}$ and the time step given by $\Delta t = 0.1 h^2$.}
    \label{fig:analytical}
\end{figure}

To verify the formulation of \texttt{CDE-1D}, the residual is evaluated on these analytical configurations for increasing spatial and temporal resolution. The \texttt{CDE-1D} residual is adapted to the low aspect ratio limit by setting $R \equiv 1$ throughout the domain of integration. Figure~\ref{fig:analytical} shows the results of this benchmark for $r_B = 1$, $\beta=0.5$, $B_{\phi,0} =1$ and $\sigma_0=1e{+}6$ and shows a clear convergence of the residual with decreasing grid cell size and time step.

\end{document}